\documentclass[twocolumn,tighten]{aastex63}
\usepackage{amsmath}
\usepackage{natbib}

\definecolor{mylinkcolor}{cmyk}{1.0,0.5,0.0,0.0}   % was {1,1,0,0}
\hypersetup{linkcolor=blue,citecolor=blue,filecolor=blue,urlcolor=blue}

\newcommand{\loggf}{\mbox{$\log(gf)$}}
\newcommand{\kmsec}{\mbox{km~s$^{\rm -1}$}}
\newcommand{\logg}{\mbox{log~{\it g}}}
\newcommand{\msun}{\mbox{$M_{\odot}$}}
\newcommand{\teff}{\mbox{$T_{\rm eff}$}}
\newcommand{\vt}{\mbox{$v_{\rm t}$}}
\newcommand{\rpro}{\mbox{{\it r}-process}}
\newcommand{\spro}{\mbox{{\it s}-process}}
\newcommand{\ipro}{\mbox{{\it i}-process}}
\newcommand{\ncap}{\mbox{{\it n}-capture}}
\newcommand{\rettwolong}{\object[NAME RETICULUM II]{Reticulum~II}}

\newcommand{\sextans}{\object[NAME SEXTANS DSPH]{Sextans}}
\newcommand{\carina}{\object[NAME CARINA DSPH]{Carina}}
\newcommand{\sculptor}{\object[NAME SCULPTOR DSPH]{Sculptor}}
\newcommand{\canvenone}{\object[NAME CANES VENATICI I DSPH]{Canes Venatici}}
\newcommand{\draco}{\object[NAME DRACO DSPH]{Draco}}
\newcommand{\ursaminor}{\object[NAME UMI DSPH]{Ursa Minor}}
\newcommand{\rone}{\mbox{\it r}-I}

\newcommand{\logeps}[1]{$\log\varepsilon$(#1)}

\newcommand{\sexcempno}{\mbox{J1008$+$0001}}

\shorttitle{Sextans dSph Stars at Large Radius}
\shortauthors{Roederer et al.}
\accepted{for publication in the Astrophysical Journal}

\begin{document}

\title{%
Abundance Analysis of Stars at Large Radius in the Sextans 
Dwarf Spheroidal Galaxy\footnote{%
This paper includes data gathered at the 6.5~meter Magellan Telescopes 
located at Las Campanas Observatory, Chile.
Other observations reported here were obtained 
at the MMT Observatory, a joint facility of the 
Smithsonian Institution and the University of Arizona.
This paper is also based on 
archival observations collected at the 
European Southern Observatory under ESO program(s) 0102.B-0786(C).
}
}

\author[0000-0001-5107-8930]{Ian U.\ Roederer}
\affiliation{%
Department of Astronomy, University of Michigan,
Ann Arbor, MI 48109, USA}
%\affiliation{%
%Department of Physics, North Carolina State University,
%Raleigh, NC 27695, USA
\affiliation{%
Joint Institute for Nuclear Astrophysics -- Center for the
Evolution of the Elements (JINA-CEE), USA}
\email{Email:\ iur@umich.edu}

\author[0000-0002-6021-8760]{Andrew B.\ Pace}
\affiliation{%
McWilliams Center for Cosmology, Carnegie Mellon University, 
Pittsburgh, PA 15213, USA}

\author[0000-0003-4479-1265]{Vinicius M.\ Placco}
\affiliation{%
NSF's NOIRLab, Tucson, AZ 85719, USA}

\author[0000-0003-2352-3202]{Nelson Caldwell}
\affiliation{%
Harvard-Smithsonian Center for Astrophysics, Cambridge, MA 02138, USA}

\author[0000-0003-2644-135X]{Sergey E. Koposov}
\affiliation{%
Institute for Astronomy, University of Edinburgh, 
Royal Observatory, Edinburgh EH9 3HJ, UK}
\affiliation{%
Institute of Astronomy, University of Cambridge, 
Cambridge CB3 0HA, UK}
\affiliation{%
Kavli Institute for Cosmology, University of Cambridge, 
Cambridge CB3 0HA, UK}

\author[0000-0002-3856-232X]{Mario Mateo}
\affiliation{%
Department of Astronomy, University of Michigan,
Ann Arbor, MI 48109, USA}

\author[0000-0002-7157-500X]{Edward W.\ Olszewski}
\affiliation{The University of Arizona, Steward Observatory, 
933 North Cherry Avenue, Tucson, AZ 85721}

\author[0000-0003-2496-1925]{Matthew G.\ Walker}
\affiliation{%
McWilliams Center for Cosmology, Carnegie Mellon University, 
Pittsburgh, PA 15213, USA}

\begin{abstract}

We present stellar parameters and chemical abundances of 
30~elements for five stars located at large radii
(3.5--10.7 times the half-light radius)
in the Sextans dwarf spheroidal galaxy.
We selected these stars using
proper motions, radial velocities, and metallicities,
and we confirm them as metal-poor members of Sextans with
$-3.34 \leq$ [Fe/H] $\leq -2.64$
using high-resolution optical spectra collected with the
Magellan Inamori Kyocera Echelle spectrograph.
Four of the five stars exhibit normal abundances of C
($-0.34 \leq$ [C/Fe] $\leq +0.36$),
mild enhancement of the $\alpha$ elements Mg, Si, Ca, and Ti
([$\alpha$/Fe] = $+0.12 \pm 0.03$),
and unremarkable abundances of
Na, Al, K, Sc, V, Cr, Mn, Co, Ni, and Zn.
We identify three chemical signatures
previously unknown among stars in Sextans.
One star
exhibits large overabundances 
([X/Fe] $> +1.2$)
of C, N, O, Na, Mg, Si, and K,
and large deficiencies of heavy elements
([Sr/Fe] = $-2.37 \pm 0.25$,
 [Ba/Fe] = $-1.45 \pm 0.20$,
 [Eu/Fe] $< +0.05$),
establishing it as a member of the class of
carbon-enhanced metal-poor stars with 
no enhancement of neutron-capture elements.
Three stars exhibit moderate enhancements of Eu
($+0.17 \leq$ [Eu/Fe] $\leq +0.70$),
and the abundance ratios among 12 neutron-capture elements
are indicative of $r$-process nucleosynthesis.
Another star is highly enhanced in Sr
relative to heavier elements
([Sr/Ba] = $+1.21 \pm 0.25$).
These chemical signatures can all be attributed to
massive, low-metallicity stars or their end states.
Our results, the first for stars at large radius in \sextans,
demonstrate that these stars were formed in chemically 
inhomogeneous regions,
such as those found in ultra-faint dwarf galaxies.

\end{abstract}

\keywords{%
Dwarf spheroidal galaxies (420);
Nucleosynthesis (1131);
Stellar abundances (1577)
}

\section{Introduction}
\label{sec:intro}

The chemical compositions of old stars 
reflect which elements were produced,
and in what amounts, 
by the earliest generations of stars and supernovae.
Old stars are found in many
Galactic environments, 
including the surviving populations of
dwarf galaxies surrounding the Milky Way.
The star-formation histories of
the lowest mass dwarf galaxies,
often referred to as
ultra-faint dwarf (UFD) galaxies
indicate that these systems
formed large fractions---up to $\approx 80$\%---of their
stars before the end of reionization
\citep{brown14}.
Stellar chemistry supports this conclusion.
Detailed chemical analysis of individual stars in UFD galaxies
reveals that they host relatively high fractions of
stars that may have formed from the remnants of 
zero-metallicity Population~III stars
(\citealt{frebel15araa}, and references therein).

More massive dwarf galaxies, 
often referred to as 
classical dwarf spheroidal (dSph) galaxies,
also formed relatively high fractions of their stars 
at early times (e.g., \citealt{revaz09,weisz14}).
The dSph galaxies are massive enough to have sustained
internal chemical evolution,
so chemical signatures 
associated with the earliest stars and supernovae
are rare
(e.g., \citealt{starkenburg10,kirby11mdf}),
but present
(e.g., \citealt{fulbright04dra,frebel10scl,skuladottir23scl}).

Most previous studies have focused on stars in the
central regions of dSph galaxies, 
but recent efforts have 
confirmed members at large separations from their centers.
These efforts have been based on
spectroscopic followup of wide-field photometric searches
(e.g., \citealt{munoz05,munoz06,westfall06,hendricks14})
or wide-field broadband photometry combined with
proper motion measurements from the
Gaia mission \citep{gaia16main}.
Studies by
\citet{chiti21,chiti23},
\citet{filion21}, 
\citet{longeard22,longeard23},
\citet{qi22},
\citet{yang22}, and
\citet{sestito23scl,sestito23umi}
have shown that several dSph and UFD galaxies
contain stars near their
tidal radii.
These extended stellar halos may have formed through 
dwarf galaxy mergers
\citep{rey19,tarumi21c}, and
multiple mergers may have occurred
within individual dSph galaxies
around the Milky Way \citep{griffen18,deason23}.
These stars frequently exhibit low metallicities, 
[Fe/H] $< -2$.
The outer regions of UFD and dSph galaxies
may host previously unrecognized 
reservoirs of stars whose chemical enrichment was potentially dominated by
the earliest generations of stars and supernovae.

Our study builds on previous work by
examining the chemistry of stars in the outer regions
of the \sextans\ dSph galaxy for the first time.
\sextans\ is 
89~kpc from the center of the Milky Way
\citep{fritz18}.
\citet{battaglia22} computed orbit integrations for \sextans\ 
that account for the reflex motion of the 
Large Magellanic Cloud on the Milky Way.
These calculations indicate that \sextans\ is
on a moderately eccentric orbit ($e \approx 0.28$),
with an orbital pericenter around 72~kpc and
an orbital apocenter around 129~kpc.
The period of star formation in \sextans\ was mainly limited to
$\approx$0.8~Gyr \citep{kirby11alpha} within
the first $\approx$1.3~Gyr after the Big Bang \citep{bettinelli18}.

\sextans\ exhibits evidence for internal stellar substructure.
\citet{kleyna04} and \citet{walker06} 
identified possible dynamically cold
substructure near the core of \sextans.
\citet{battaglia11} found evidence for 
two chemodynamical stellar populations in \sextans.
\citet{roderick16} found evidence of an extended, 
gravitationally bound stellar structure within
the tidal radius.
% ($83\farcm2 \pm 7\farcm1$; $2.08 \pm 0.18$~kpc).
This stellar substructure is probably unrelated to
disruptive tidal effects,
as \citet{cicuendez18a} found
no significant distortions or signs of 
tidal disturbances in \sextans.
The stellar substructure could be related to accretion.
\citet{cicuendez18b} identified a ring-like structure
surrounding the inner regions
($\approx 15$--20$^{\prime}$)
of \sextans.
This feature is characterized by a 
small velocity offset and lower metallicity
relative to the surrounding stellar fields \citep{walker09a}.
Finally, \citet{kim19} identified a metal-poor stellar overdensity
in \sextans\ that might be a 
low-mass star cluster undergoing dissolution.
\sextans\ is not unusual among dSph galaxies
in exhibiting substructure
(e.g., \citealt{olszewski85,battaglia06,olszewski06,amorisco14,pace20}).

Previous studies have derived detailed
chemical abundances of stars in \sextans\
using high-resolution spectroscopy
\citep{shetrone01,aoki09sex,tafelmeyer10,honda11sex,
aoki20, % note that this is a different Aoki
lucchesi20,theler20,mashonkina22,fernandes23}.
These studies
have been limited to stars near the center of \sextans, 
within the inner $\approx40^{\prime}$ or so.
They have found chemical abundance behaviors that are
relatively typical for dSph galaxies.
These signatures include
enhanced abundances of $\alpha$ elements
(where $\alpha$ represents O, Mg, Si, Ca, and Ti)
in the lowest metallicity stars
([Fe/H] $< -2.8$ in \sextans).
This behavior indicates that
core-collapse supernovae
dominated the chemical enrichment at early times
when the most metal-poor stars likely were forming.
The [$\alpha$/Fe] ratios exhibit a so-called ``knee''
when plotted against [Fe/H],
either at [Fe/H] $\approx -2.5$ or $-2.0$.
Stars with metallicities higher than this knee
exhibit lower [$\alpha$/Fe] ratios,
a behavior typically explained by contributions
from Type~Ia supernovae.
Two knees could indicate
the presence of slightly older and slightly
younger populations of stars,
which could be a potential accretion signature
\citep{benitezllambay16,reichert20,mashonkina22}.
The most metal-poor stars in \sextans\
exhibit subsolar [Sr/Fe] and [Ba/Fe] ratios,
which might signal the presence of
small amounts of material produced by the 
weak component of the rapid neutron-capture process (\rpro).
Some metal-rich 
([Fe/H] $> -2.2$)
stars in \sextans\
exhibit signatures of the slow neutron-capture process (\spro),
which appears on delayed timescales and occurs in
low- or intermediate-mass stars that pass through the 
asymptotic giant branch (AGB) phase of evolution.
Few carbon-enhanced stars are known in \sextans\
\citep{honda11sex,theler20,mashonkina22}.

We report on the chemical abundances of five stars 
at large radius in \sextans.
These stars exhibit abundance patterns previously unrecognized in \sextans,
including large enhancements of carbon and other light elements,
and several distinct signatures among the heaviest elements.
Our manuscript is structured as follows.
Section~\ref{sec:data} presents our target selection and 
new spectroscopic data.
Section~\ref{sec:analysis} describes our abundance analysis
of these spectra.
Section~\ref{sec:abundresults} presents our results
and compares them with previous work.
Section~\ref{sec:abundsummary} discusses these results,
and 
Section~\ref{sec:conclusions} summarizes our conclusions.

\section{Data}
\label{sec:data}

\subsection{Target Selection}
\label{sec:targets}

Our targets were selected as
confirmed members in radial velocity surveys (Pace et al., in preparation) 
or from a proper-motion-based selection 
\citep{pace22} 
using Gaia's early data release 3 (EDR3; \citealt{gaia21edr3}).
We focused on bright ($G \lesssim 17.5$)
and distant ($R_{\rm e}/R_{\rm h}\gtrsim3$) stars,
where 
$G$ is the Gaia broadband photometric magnitude,
$R_{\rm e} \equiv \sqrt{x^2 + y^2/q^2}$ 
is the deprojected elliptical radius, and
$R_{\rm h}$ is the \sextans\ half-light radius
($16\farcm9 \pm 0\farcm1$; \citealt{munoz18}).
We identified \mbox{J1015$-$0238} 
as a radial velocity member from 
spectra collected using the Hectochelle spectrograph \citep{szentgyorgyi11}
at the MMT Observatory.
We identified \mbox{J1018$-$0209} and \mbox{J1008$+$0001}
from archival spectra collected using the 
Fibre Large Array Multi Element Spectrograph's GIRAFFE instrument
\citep{pasquini02}
at the Very Large Telescope. 
Other targets lack previous radial velocity measurements, so
we considered their membership probabilities from \citeauthor{pace22}\
and examined photometry from the ninth data release of 
the Dark Energy Camera Legacy Survey (DECaLS DR9; \citealt{dey19}).
We compared the locations of candidate members
and spectroscopic members 
in $g-r$ versus $g$ color-magnitude diagrams 
and $g-r$ versus $r-z$ color-color diagrams.
We obtained low signal-to-noise (S/N) spectra (Section~\ref{sec:mike})
to measure radial velocities to confirm membership 
before obtaining longer observations with higher S/N ratios.
Table~\ref{tab:targets} lists the
target names, coordinates, 
the ratio of $R_{\rm e}$ to $R_{\rm h}$,
selected photometry, and reddening estimates
for the stars in our sample.

\begin{deluxetable*}{ccccccccccccc}
\tablecaption{Star Names, Coordinates, Photometry, and Reddening
\label{tab:targets}}
\rotate
\tablewidth{0pt}
\tabletypesize{\scriptsize}
\tablehead{
\colhead{Source\_ID} &
\colhead{Star Name} &
\colhead{Star Name} &
\colhead{R.A.} &
\colhead{Dec.} &
\colhead{$R_{\rm e}/R_{\rm h}$} &
\colhead{$G$} &
\colhead{$g$} &
\colhead{$B$} & 
\colhead{$V$} & 
\colhead{$E(B-V)$} &
\colhead{$E(B-V)$} &
\colhead{$E(B-V)$} \\
\colhead{(Gaia\tablenotemark{a})} &
\colhead{(SDSS\tablenotemark{b})} &
\colhead{(adopted)} &
\colhead{(J2000)} &
\colhead{(J2000)} &
\colhead{} &
\colhead{(Gaia)} &
\colhead{(SDSS)} &
\colhead{(\tablenotemark{c})} & 
\colhead{(\tablenotemark{c})} & 
\colhead{(SF11\tablenotemark{d})} &
\colhead{(Na~\textsc{i})} &
\colhead{(adopted)}
}
\startdata
\multicolumn{13}{c}{Stars with high-S/N observations} \\
\hline
3831812247731524608 & J100801.54$+$000108.1 & J1008$+$0001 & 10:08:01.54 & $+$00:01:08.1 & 10.68 & 18.49 & 18.80 & 19.38 & 18.27 & 0.027 &    0.012 & 0.02 \\
3828963348679468032 & J101039.85$-$022007.8 & J1010$-$0220 & 10:10:39.85 & $-$02:20:07.8 &  3.56 & 17.23 & 18.12 & 18.68 & 17.63 & 0.034 & $>$0.044 & 0.04 \\
3828784987277714560 & J101542.20$-$023838.6 & J1015$-$0238 & 10:15:42.21 & $-$02:38:38.7 &  6.36 & 17.08 & 18.01 & 18.58 & 17.50 & 0.031 &    0.021 & 0.03 \\
3830390720930784640 & J101800.19$-$015521.4 & J1018$-$0155 & 10:18:00.20 & $-$01:55:21.5 &  5.79 & 16.98 & 18.03 & 18.63 & 17.47 & 0.043 & $>$0.066 & 0.05 \\
3830319390113933952 & J101837.07$-$020936.2 & J1018$-$0209 & 10:18:37.08 & $-$02:09:36.3 &  7.06 & 17.55 & 17.91 & 18.52 & 17.33 & 0.038 & $>$0.031 & 0.04 \\
\hline
\multicolumn{13}{c}{Stars with low-S/N observations} \\
\hline
3829054779943345536 & J101341.76$-$021124.4 & J1013$-$0211 & 10:13:41.76 & $-$02:11:24.4 &  3.04 & 16.75 & 17.86 & 18.48 & 17.27 & 0.033 & \nodata  & \nodata \\
3830721875794075904 & J101435.84$-$005401.4 & J1014$-$0054 & 10:14:35.84 & $-$00:54:01.4 &  3.28 & 16.84 & 17.95 & 18.58 & 17.34 & 0.034 & \nodata  & \nodata \\
\enddata      
%\tablecomments{%
%}
\tablenotetext{a}{%
%Early data release 3 (EDR3; \citealt{gaia21edr3})
Gaia EDR3 \citep{gaia21edr3}
}
\tablenotetext{b}{%
Sloan Digital Sky Survey data release 13 (SDSS DR13; \citealt{albareti17})
}
\tablenotetext{c}{%
The $B$ and $V$ magnitudes are calculated from the 
SDSS $g$ magnitude using 
the Population~II star transformations of \citet{jordi06}.
}
\tablenotetext{d}{%
\citet{schlafly11}
}
\end{deluxetable*}

\begin{figure}
\begin{center}
\includegraphics[angle=0,width=3.4in]{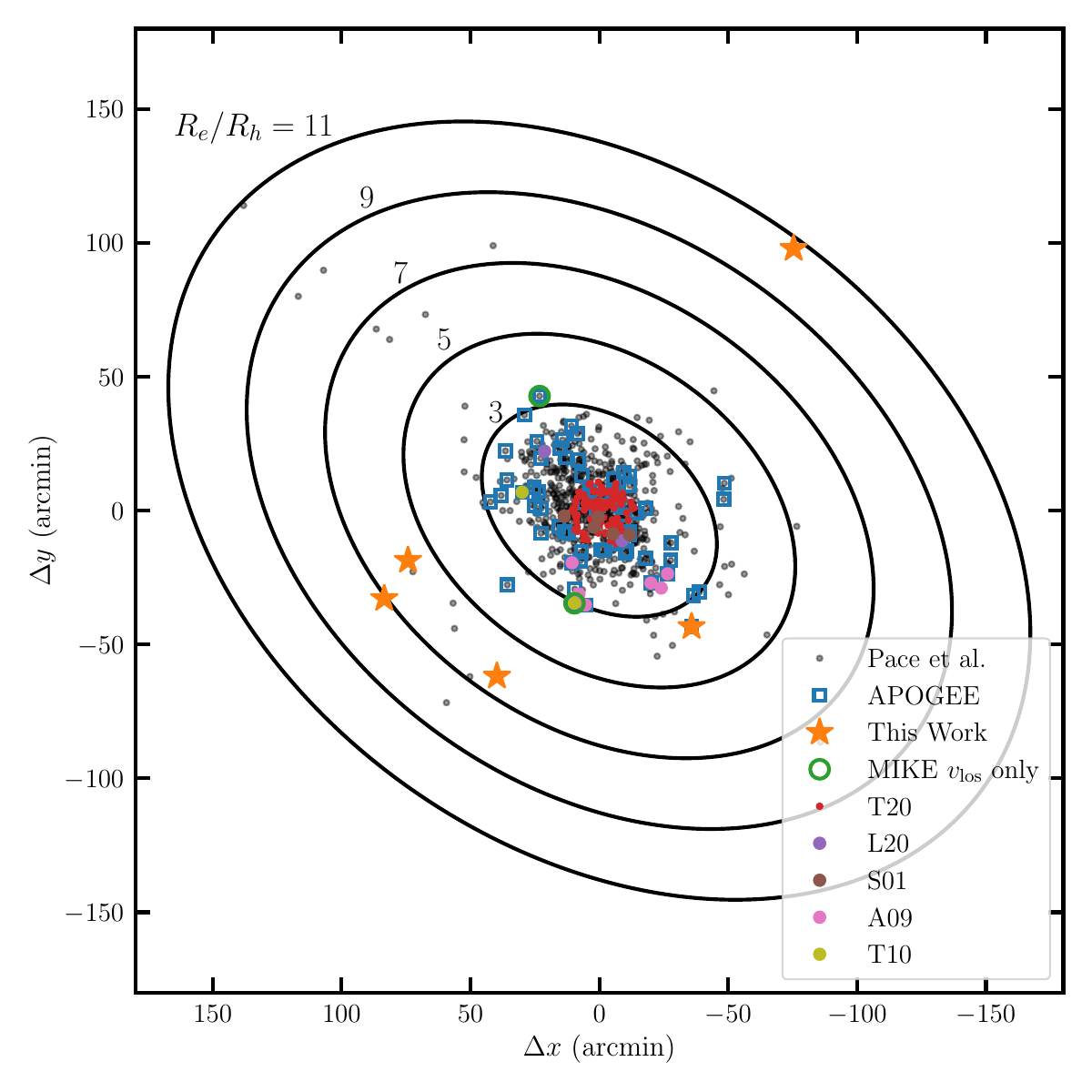}
\end{center}
\caption{
\label{fig:skyplot}
Plot of the spatial distribution of 
our sample (orange stars and green circles)
and previous spectroscopic samples of 
stars in \sextans:\
T20 = \citet{theler20},
L20 = \citet{lucchesi20},
S01 = \citet{shetrone01},
A09 = \citet{aoki09sex}, and
T10 = \citet{tafelmeyer10}.
The small gray dots mark stars observed 
in our medium-resolution work (Pace et al., in preparation).
The ellipses indicate multiples of $R_{\rm h}$.
}
\end{figure}

Figure~\ref{fig:skyplot} illustrates the spatial distribution 
of the stars in our sample and previous 
high-resolution and medium-resolution spectroscopic samples.
The stars in our high-S/N sample (Section~\ref{sec:mike}), 
shown by the orange stars, span 
$3.5 < R_{\rm e} / R_{\rm h} < 10.7$.
These stars are located at much larger radii 
than previous high-resolution samples,
which are concentrated within 4 $R_{\rm e}$/$R_{\rm h}$,
and the vast majority of which are within 2 $R_{\rm e}$/$R_{\rm h}$.
The King tidal (or limiting) radius,
$R_{\rm t}$, 
is uncertain for \sextans,
with estimates of
3.7~$R_{\rm h}$ \citep{munoz18},
5.0~$R_{\rm h}$ \citep{roderick16}, and
6.2~$R_{\rm h}$ \citep{tokiwa23}.
At least two, and possibly four,
of the five stars in our high-S/N sample 
are beyond $R_{\rm t}$, which is
roughly the radius at which the stellar overdensity
of the dwarf galaxy falls below that of the 
Milky Way foreground.

\begin{figure}
\begin{center}
\includegraphics[angle=0,width=3.4in]{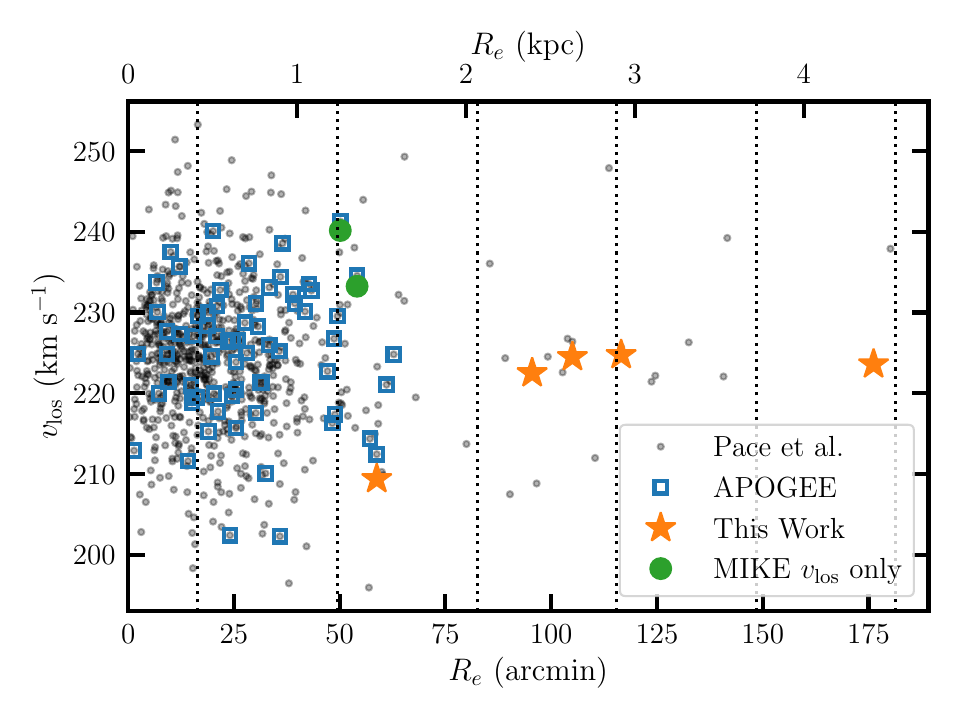}
\end{center}
\caption{
\label{fig:vlos}
Comparison of $R_{\rm e}$
versus $v_{\rm los}$
for stars in \sextans.
The scale on the top axis assumes a distance of
86.1~kpc \citep{helmi18}.
Vertical dotted lines mark 1, 3, 5, 7 9, and 11 times $R_{\rm h}$.
The five stars observed with high S/N in the present study are
marked with orange stars, and the two stars observed
with low S/N in the present study are marked with green circles.
Previous samples from APOGEE \citep{abdurrouf22} and our own 
medium-resolution work (Pace et al., in preparation) are marked with
open blue squares and small gray dots, respectively.
}
\end{figure}

Figure~\ref{fig:vlos} illustrates the 
line-of-sight velocity, $v_{\rm los}$, 
as a function of radial distance from the center of the \sextans\ dSph.
Our $v_{\rm los}$ measurements agree with previous values, when 
available, 
and they cluster around the systemic $v_{\rm los}$ 
of the \sextans\ dSph, $224.3 \pm 0.1$~\kmsec\ \citep{walker09c}.
The stars in our sample
are high-probability members of \sextans.

\subsection{Observations}
\label{sec:mike}

We used the
Magellan Inamori Kyocera Echelle (MIKE; \citealt{bernstein03}) spectrograph
on the Landon Clay (Magellan~II) Telescope at 
Las Campanas Observatory, Chile,
to collect high-resolution spectra of seven stars in \sextans.
These spectra were obtained on several nights in 2021 and 2022
during dark time and under excellent seeing conditions
($\approx 0\farcs4$--0\farcs8).
The 0\farcs7$\times$5\farcs0 
entrance slit and 2$\times$2 binning on the CCD yield a 
spectral resolving power of 
$R \equiv \lambda/\Delta\lambda \sim 41,000$ on the blue spectrograph
(3350 $< \lambda <$ 5000~\AA) and $R \sim 36,000$ on the red spectrograph
(5000 $< \lambda <$ 9150~\AA).
We observed each star using
a series of exposures, ranging from 1500~s to 2300~s each.
We obtained ThAr comparison spectra
immediately before or after the series of exposures
of each star.
Table~\ref{tab:mike} summarizes the
observing date, UT at mid observation, total exposure time, 
heliocentric $v_{\rm los}$,
and S/N ratios at several 
wavelengths in the co-added spectrum
of each star.
We focus our attention 
on the five stars with high S/N ratios.

\begin{deluxetable*}{ccccccccc}
\tablecaption{Log of MIKE Observations
\label{tab:mike}}
\tablewidth{0pt}
\tabletypesize{\scriptsize}
\tablehead{
\colhead{Star name} &
\colhead{Obs.\ date} &
\colhead{UT} &
\colhead{$t_{\rm exp}$} &
\colhead{$v_{\rm los}$} &
\colhead{S/N@3950~\AA} &
\colhead{S/N@4550~\AA} &
\colhead{S/N@5200~\AA} &
\colhead{S/N@6700~\AA} \\
\colhead{} &
\colhead{} &
\colhead{} &
\colhead{(hr)} &
\colhead{(\kmsec)} &
\colhead{(pix$^{-1}$)} &
\colhead{(pix$^{-1}$)} &
\colhead{(pix$^{-1}$)} &
\colhead{(pix$^{-1}$)}
}
\startdata
\multicolumn{9}{c}{Stars with high-S/N observations} \\
\hline
J1008$+$0001 & 2022/03/03 & 04:42 & 5.56 & $+$223.6 & 13 & 30 & 26 & 61 \\
J1010$-$0220 & 2021/01/12 & 08:22 & 1.11 & $+$209.0 & 17 & 34 & 30 & 67 \\
             & 2021/01/13 & 04:58 & 2.56 & $+$209.8 &    &    &    &    \\
J1015$-$0238 & 2021/01/12 & 06:18 & 2.89 & $+$224.5 & 15 & 31 & 28 & 64 \\
J1018$-$0155 & 2021/01/13 & 07:34 & 2.47 & $+$222.5 & 16 & 33 & 30 & 70 \\
J1018$-$0209 & 2021/12/05 & 07:23 & 1.61 & $+$224.7 & 13 & 30 & 27 & 66 \\
             & 2021/12/06 & 07:24 & 1.67 & $+$224.8 &    &    &    &    \\
\hline
\multicolumn{9}{c}{Stars with low-S/N observations} \\
\hline
J1013$-$0211 & 2021/01/12 & 03:59 & 0.19 & $+$242.1 &  3 &  8 &  8 & 19 \\
J1014$-$0054 & 2021/01/12 & 04:27 & 0.19 & $+$234.1 &  3 &  8 &  7 & 18 \\
\enddata      
%\tablecomments{%
%}
%\tablenotetext{a}{}
\end{deluxetable*}

We use the CarPy MIKE reduction pipeline
\citep{kelson00,kelson03}
to perform the overscan subtraction, pixel-to-pixel flat field division,
image coaddition, cosmic ray removal, sky and scattered-light subtraction,
rectification of the tilted slit profiles along the orders,
spectrum extraction, and wavelength calibration.
We use the IRAF \citep{tody93} software package
to stitch together and
continuum-normalize the spectra.

\begin{figure*}
\begin{center}
\includegraphics[angle=0,width=7.0in]{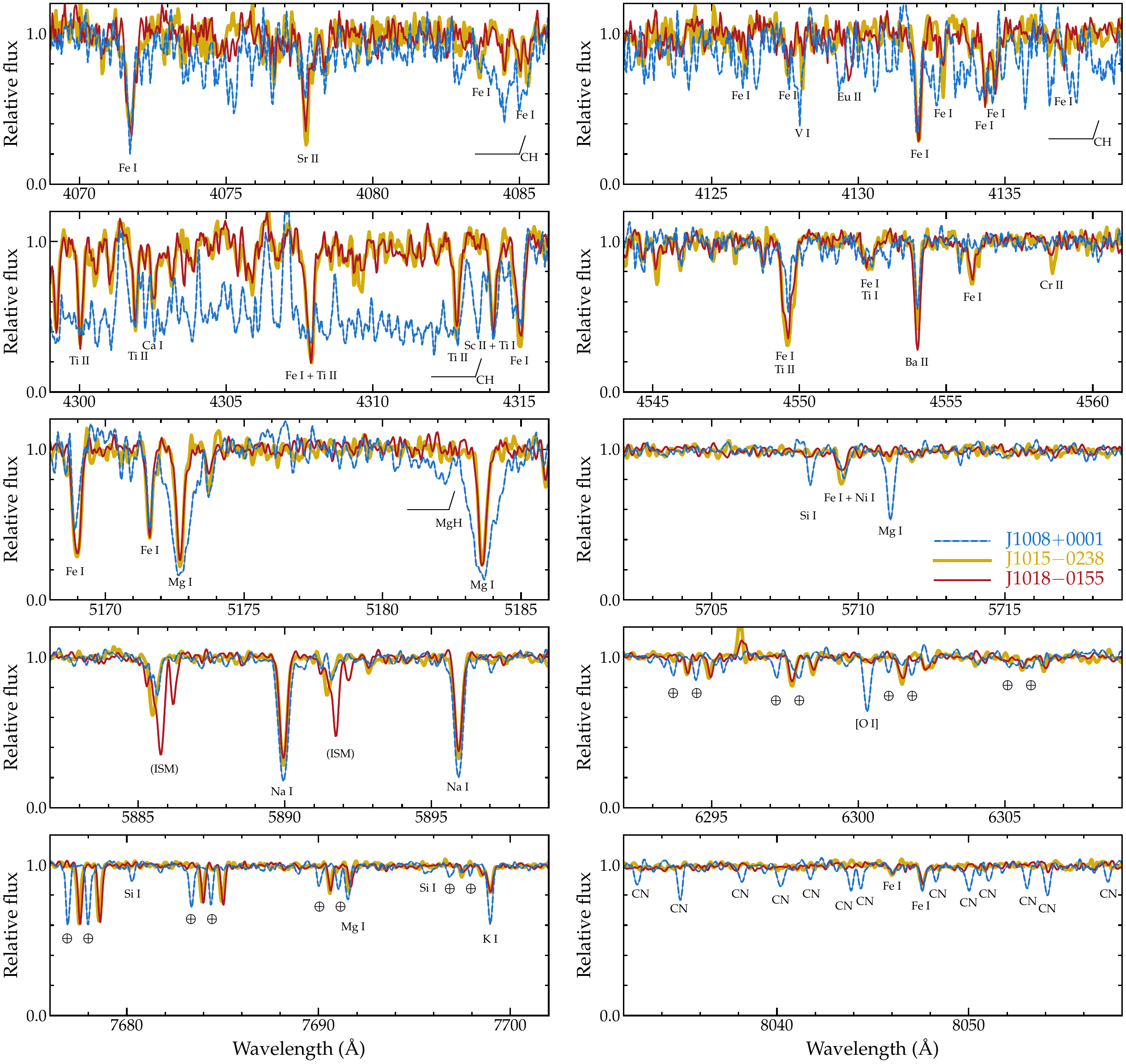}
\end{center}
\caption{
\label{fig:plotspec}
Selected regions of the MIKE spectra of three stars.
Several absorption lines are identified.
These three stars have similar \teff\ and \logg,
so the differences in the line strengths mainly reflect
abundance differences.
Several interstellar medium (ISM) and O$_{2}$ telluric lines ($\Earth$)
are detected and marked.
The telluric lines shift in velocity relative to the stellar lines, and
they are marked at their approximate wavelengths 
in \sexcempno.
}
\end{figure*}

Figure~\ref{fig:plotspec} illustrates several regions
of the spectra around lines of interest.
A few key features are immediately discernible.
First, the differences in line strengths are mainly due to 
differences in abundance, because these stars
have similar stellar parameters (Section~\ref{sec:modelatm}).
Secondly, lines of Ti and Fe exhibit only minimal differences,
indicating that these stars have similar metallicities
to within a factor of a few (Section~\ref{sec:modelatm}).
Thirdly,
one star, \sexcempno,
has much stronger CH, CN, [O~\textsc{i}], Na~\textsc{i}, 
Mg~\textsc{i}, Si~\textsc{i}, and K~\textsc{i} lines,
while its Ba~\textsc{ii} and Eu~\textsc{ii} lines
are much weaker than those in other stars
(Section~\ref{sec:cempno}).
Finally,
\mbox{J1018$-$0155}
(along with \mbox{J1010$-$0220} and \mbox{J1018$-$0209}; not shown)
exhibits moderately strong Eu~\textsc{ii} lines,
suggesting that these three stars
are enhanced in \rpro\ elements (Section~\ref{sec:heavy}).

We measure $v_{\rm los}$ by cross-correlating the
echelle order containing the Mg~\textsc{i} \textit{b} triplet 
against a metal-poor template spectrum obtained with MIKE,
using the IRAF ``fxcor'' task.
We calculate the heliocentric velocity corrections using the 
IRAF ``rvcorrect'' task.
\citet{roederer14c} estimated uncertainties of $\approx 0.7$~\kmsec\
for $v_{\rm los}$ values measured by this method.
Repeat observations of 
\mbox{J1010$-$0220} and \mbox{J1018$-$0209} 
yield consistent $v_{\rm los}$ values
that support this estimate.

\section{Analysis}
\label{sec:analysis}

We describe our derivation of stellar parameters
(Table~\ref{tab:atm}) and abundances
(Tables~\ref{tab:lineabund} through \ref{tab:abund3}) in this section.
We define the abundance of element X as
\logeps{X}~$\equiv \log_{10}(N_{\rm X}/N_{\rm H})+$12.0,
where $N_{\rm X}$ represents the number density of element X.
We define the abundance ratio of X and Fe relative to the
solar ratio as
[X/Fe] $\equiv \log_{10} (N_{\rm X}/N_{\rm Fe}) - \log_{10} (N_{\rm X}/N_{\rm Fe})_{\odot}$.
We adopt the solar abundances,
listed in Table~\ref{tab:abund1},
from \citet{asplund09}.
By convention, abundances or ratios denoted 
with the ionization state
(e.g., [Fe~\textsc{ii}/H])
are understood to be 
the total elemental abundance as derived from transitions of that
particular ionization state 
after \citet{saha21} ionization corrections have been applied.

\begin{deluxetable}{cccccc}
\tablecaption{Model Atmosphere Parameters
\label{tab:atm}}
\tablewidth{0pt}
\tabletypesize{\scriptsize}
\tablehead{
\colhead{Star name} &
\colhead{\teff} &
\colhead{\logg} &
\colhead{\vt} &
\colhead{[M/H]\tablenotemark{a}} &
\colhead{[Fe~\textsc{i}/H]\tablenotemark{b}} \\
\colhead{} &
\colhead{(K)} &
\colhead{} &
\colhead{(\kmsec)} &
\colhead{} &
\colhead{}
}
\startdata
J1008$+$0001 & 4405 & 1.07 & 2.25 & $-$3.43 & $-$2.97 \\
J1010$-$0220 & 4405 & 0.79 & 2.15 & $-$3.03 & $-$3.34 \\
J1015$-$0238 & 4441 & 0.79 & 2.35 & $-$2.73 & $-$2.64 \\
J1018$-$0155 & 4423 & 0.72 & 2.45 & $-$2.89 & $-$2.81 \\
J1018$-$0209 & 4396 & 0.67 & 2.45 & $-$2.86 & $-$2.75 
\enddata      
%\tablecomments{%
%}
\tablenotetext{a}{%
[M/H] $\equiv$ [Fe~\textsc{ii}/H]}
\tablenotetext{b}{%
Includes NLTE correction}
\end{deluxetable}

\subsection{Model Atmospheres}
\label{sec:modelatm}

We derive model atmosphere parameters using 
a combination of quantities measured from the
spectra themselves
and values adopted from external catalogs.
We interpolate models from the 1D 
ATLAS9 grid of $\alpha$-enhanced models \citep{castelli04}
using an interpolation code provided by
A.\ McWilliam (2009, private communication).  

We rely on abundances derived from 
equivalent widths (EWs) of 
Fe~\textsc{i} and \textsc{ii} lines
as part of this process.
We measure EWs
using a semi-automated 
routine that fits Voigt or Gaussian line profiles to 
continuum-normalized spectra
\citep{roederer14c}.
Each line is inspected visually.
A telluric spectrum is simultaneously compared with the stellar spectrum,
and we discard
any lines that appear to be contaminated by
telluric absorption.
These Fe~\textsc{i} and \textsc{ii} lines are listed in 
Table~\ref{tab:lineabund}.
We derive Fe abundances 
using a recent version of the 
line analysis software MOOG
(\citealt{sneden73,sobeck11}; 2017 version), which
assumes local thermodynamic equilibrium (LTE).~
We adopt damping constants for collisional broadening
with neutral hydrogen from \citet{barklem00h}
and \citet{barklem05feii}, when available,
otherwise
we adopt the standard \citet{unsold55} recipe.
We discard strong Fe lines with $\log$(EW/$\lambda$) $> -4.5$.
The weakest lines employed in our analysis
have EW~$\approx 7$~m\AA\ (Table~\ref{tab:lineabund}).

\startlongtable
\begin{deluxetable*}{cccccccccccccccc}
\tablecaption{Line Atomic Data and Derived Abundances
\label{tab:lineabund}}
\tablewidth{0pt}
\tabletypesize{\scriptsize} %\tiny}
\tablehead{
\colhead{} &
\colhead{} &
\colhead{} &
\colhead{} &
\colhead{} &
\colhead{} &
\multicolumn{4}{c}{J1008$+$0001} &
\colhead{} &
\multicolumn{4}{c}{J1010$-$0220} &
\colhead{\ldots} \\
\cline{7-10}
\cline{12-15}
\colhead{Species} &
\colhead{$\lambda$} &
\colhead{E.P.} &
\colhead{\loggf} &
\colhead{\loggf} &
\colhead{\loggf} &
\colhead{EW} &
\colhead{U.L.} &
\colhead{$\log\varepsilon$} &
\colhead{NLTE} &
\colhead{} &
\colhead{EW} &
\colhead{U.L.} &
\colhead{$\log\varepsilon$} &
\colhead{NLTE} &
\colhead{\ldots} \\
\colhead{} &
\colhead{} &
\colhead{} &
\colhead{} &
\colhead{unc.} &
\colhead{ref.} &
\colhead{} &
\colhead{flag} &
\colhead{(LTE)} &
\colhead{cor.} &
\colhead{} &
\colhead{} &
\colhead{flag} &
\colhead{(LTE)} &
\colhead{cor.} &
\colhead{\ldots} \\
\colhead{} &
\colhead{(\AA)} &
\colhead{(eV)} &
\colhead{} &
\colhead{} &
\colhead{} &
\colhead{(m\AA)} &
\colhead{} &
\colhead{} &
\colhead{} &
\colhead{} &
\colhead{(m\AA)} &
\colhead{} &
\colhead{} &
\colhead{} &
\colhead{\ldots} 
}
\startdata
%                                         J1008: EW,  ul,   logeps,   NLTEcor  J1010: EW,  ul,   logeps,   NLTEcor  
Li I  & 6707.80 & 0.00 &    0.17 & 0.01 & 1  &         & $<$ &  0.10   & +0.15   &&         & $<$ &  0.40   & +0.15   & \ldots \\
O I   & 6300.30 & 0.00 & $-$9.82 & 0.03 & 2  & \nodata &     &  8.07   & \nodata && \nodata & $<$ &  6.70   & \nodata & \ldots \\
O I   & 6363.78 & 0.02 &$-$10.26 & 0.03 & 2  & \nodata &     &  8.08   & \nodata && \nodata &     & \nodata & \nodata & \ldots \\
Na I  & 5682.63 & 2.10 & $-$0.71 & 0.01 & 2  &  52.2   &     &  4.96   & -0.13   &&         &     &         &         & \ldots \\
Na I  & 5688.19 & 2.10 & $-$0.41 & 0.01 & 2  &  69.9   &     &  4.91   & -0.16   &&         &     &         &         & \ldots \\
%Na I  & 5889.95 & 0.00 &    0.11 & 0.01 & 2  &         &     &         &         &&         &     &  2.73   & -0.15   & \ldots \\
%Na I  & 5895.92 & 0.00 & $-$0.19 & 0.01 & 2  &         &     &         &         &&         &     &  2.70   & -0.07   & \ldots \\
%Na I  & 6154.23 & 2.10 & $-$1.55 & 0.01 & 2  &  14.0   &     &  5.02   & -0.08   &&         &     &         &         & \ldots \\
%Na I  & 6160.75 & 2.10 & $-$1.25 & 0.01 & 2  &  24.0   &     &  5.00   & -0.08   &&         &     &         &         & \ldots \\
\vdots&\vdots  &\vdots&\vdots &\vdots&\vdots&\vdots &\vdots&\vdots  &\vdots   && \vdots  &\vdots&\vdots  & \vdots  & \vdots 
\enddata      
\tablereferences{%
1:\ \citet{smith98}, using HFS from \citet{kurucz11};
2:\ \citet{kramida21};
3:\ \citet{pehlivanrhodin17};
4:\ \citet{kramida21}, using HFS from VALD3 \citep{piskunov95,pakhomov19};
5:\ \citet{denhartog23};
6:\ \citet{denhartog21ca};
7:\ \citet{lawler89}, using HFS from \citet{kurucz11};
8:\ \citet{lawler13};
9:\ \citet{pickering01}, using corrections given in \citet{pickering02};
10:\ \citet{wood13};
11:\ \citet{lawler14}, including HFS;
12:\ \citet{wood14v}, including HFS;
13:\ \citet{sobeck07};
14:\ \citet{lawler17};
15:\ \citet{denhartog11}, including HFS;
16:\ \citet{obrian91};
17:\ \citet{denhartog14};
18:\ \citet{ruffoni14};
19:\ \citet{belmonte17};
20:\ \citet{blackwell82fe};
21:\ \citet{melendez09fe};
22:\ \citet{denhartog19};
23:\ \citet{lawler15}, including HFS;
24:\ \citet{wood14ni};
25:\ \citet{roederer12b};
26:\ \citet{biemont11};
27:\ \citet{ljung06};
28:\ \citet{kramida21}, using HFS/IS from \citet{mcwilliam98} 
        or other sources when available;
29:\ \citet{lawler01la}, using HFS from \citet{ivans06} when available;
30:\ \citet{lawler09};
31:\ \citet{li07}, using HFS from \citet{sneden09};
32:\ \citet{denhartog03};
33:\ \citet{lawler06}, using HFS/IS from \citet{roederer08a};
34:\ \citet{lawler01eu}, using HFS/IS from \citet{ivans06};
35:\ \citet{wickliffe00};
36:\ \citet{biemont00}, using HFS/IS from \citet{roederer12d}.
}
\tablecomments{%
The complete version of Table~\ref{tab:lineabund} is available 
in machine-readable form in the online edition of the journal.
A small section is shown here to illustrate its form and content.
}
%\tablenotetext{a}{%
%}
\end{deluxetable*}

Stellar effective temperatures (\teff) may be derived from 
photometric or spectroscopic methods.
We derive \teff\ values using the spectroscopic
excitation balance method,
and we apply a separate calibration \citep{frebel13} 
to transform this scale, which is generally considered to be too cool,
to the warmer photometric one.
We begin by
identifying the \teff, 
log of the surface gravity (\logg; cm~s$^{-2}$ in cgs units),
microturbulent velocity parameter (\vt), and
model metallicity ([M/H]) that
meet the following set of requirements.
We set \teff\ by
requiring no trend between the
abundance derived from Fe~\textsc{i} lines
and the lower excitation potential of each transition.
We set \vt\ by
requiring no trend between the
abundance derived from Fe~\textsc{i} lines
and the line strength.
We set \logg\ by
requiring that the mean abundances calculated 
from Fe~\textsc{i} and \textsc{ii} lines
agree within their uncertainties;
in practice, these two quantities
are closest at the edge of the model atmosphere grid at \logg\ = 0.0.
We set [M/H] by
matching the Fe abundance (from Fe~\textsc{i} lines)
plus 0.25~dex as recommended by \citeauthor{frebel13} %.
Once these values converge, 
we calculate a corrected \teff\ by extrapolating
Equation~1 of \citeauthor{frebel13} %.
The corrected \teff\ values are $\approx 250$~K warmer than the
purely spectroscopic ones for these stars.

We use the corrected \teff\ to calculate a new
\logg\ from fundamental relations:\
\begin{eqnarray}
\label{eqn:logg}
\log g = 4 \log \teff + \log(M/\msun) - 10.61 + 0.4(BC_{V}
  \nonumber\\
  + m_{V} - 5\log d + 5 - 3.1 E(B-V) - M_{\rm bol,\odot}).
\end{eqnarray}
Here, $M$ is the mass of the star, which we
assume to be 0.8~$\pm$~0.08~\msun.
$BC_{V}$ is the bolometric correction in the $V$ band
\citep{casagrande14c}.
$m_{V}$ is the apparent $V$ magnitude.
$d$ is the distance in pc, which is
assumed to be 86.1 $\pm$~2.6~kpc
\citep{helmi18}.
We rederive \vt\ and metallicity
and iterate on the stellar parameters, 
including $BC_{V}$, until the
[M/H] matches [Fe~\textsc{ii}/H] and 
there is no trend between the
abundance derived from Fe~\textsc{i} lines
and the line strength.

Equation~\ref{eqn:logg} requires an estimate of the 
reddening along the line of sight to each star, $E(B-V)$.
We estimate $E(B-V)$ by two methods.
We interpolate the dust maps presented by \citet{schlafly11}, 
which provide the $E(B-V)$ values along the line of sight,
and we assume that all of the interstellar reddening lies
in front of \sextans.
We also estimate $E(B-V)$ using the
interstellar Na~\textsc{i} \textit{D} absorption 
\citep{bohlin78,spitzer78,ferlet85},
as described in \citet{roederer18b}.
We measure the EWs by direct integration using the
IRAF ``splot'' task.
For stars \mbox{J1015$-$0238} and \sexcempno,
the ratio of the EWs of the two components of the doublet 
is $\approx$2:1 (120:65~m\AA\ and 70:35~m\AA, respectively),
the same as the ratio of the $f$-values of these transitions.
These lines are on the linear part of the curve of growth
and thus sensitive to the reddening.
For the other three stars, multiple components are present, 
the EWs are larger, and they are not in 2:1 ratios
(\mbox{J1010$-$0220}, 220:160~m\AA;
 \mbox{J1018$-$0155}, 345:235~m\AA;
 \mbox{J1018$-$0209}, 175:100~m\AA).
They are saturated and so
only yield limits on the amount of interstellar absorption.
The empirical relations between Na~\textsc{i} absorption,
$N$(H~\textsc{i} + H$_{2}$), and $E(B-V)$ 
have intrinsic scatter that corresponds to a few hundredths of a mag
in $E(B-V)$.
The two methods yield reasonably similar $E(B-V)$ values,
which we list along with our adopted averages
in Table~\ref{tab:targets}.
Our adopted set of stellar parameters is listed in
Table~\ref{tab:atm}.

We estimate the mean and uncertainty in each stellar parameter as follows.
\citet{frebel13} estimate uncertainties in \teff\ of $\approx 150$~K
using their method.
For \logg, we draw $10^{4}$ 
samples from each input parameter in the \logg\ calculation,
assuming Gaussian uncertainties.
The statistical uncertainty associated with this method is
$\approx 0.09$~dex.
The systematic uncertainty is certainly larger, 
$\sim 0.25$~dex or so \citep{jofre19}.
For a given \teff\ and \logg, 
the uncertainty in \vt\ is $\approx 0.2$~\kmsec\ and 
the uncertainty in [M/H] is $\approx 0.2$~dex.

The LTE [Fe/H] ratios derived from Fe~\textsc{i} and Fe~\textsc{ii} lines
are not forced into agreement using this method.
Non-LTE (NLTE) overionization
of neutral Fe causes the Fe abundance from Fe~\textsc{i} lines
to be underestimated \citep{thevenin99}.
NLTE corrections for Fe~\textsc{ii} lines are generally negligible.
NLTE corrections
are available for $\approx 25$ of the Fe~\textsc{i} lines
for which we have measured EWs.
We evaluate these corrections 
by interpolating the pre-computed grids
presented in the INSPECT database 
\citep{bergemann12,lind12}.
The NLTE corrections range from $+$0.10 to $+$0.14 for these five stars.
[Fe~\textsc{ii}/Fe~\textsc{i}] ionization equilibrium
is achieved within 1.8$\sigma$ after including these NLTE corrections.
We adopt the NLTE-corrected Fe abundance from Fe~\textsc{i} lines
when constructing abundance ratios of various elements relative to Fe
(i.e., [X/Fe]).

\begin{deluxetable*}{cccccccccccccc}
\tablecaption{Derived Abundances (Part 1 of 3)
\label{tab:abund1}}
\tablewidth{0pt}
\tabletypesize{\scriptsize}
\tablehead{
\colhead{} &
\colhead{} &
\colhead{} &
\multicolumn{5}{c}{J1008$+$0001} &
\colhead{} &
\multicolumn{5}{c}{J1010$-$0220} \\
\cline{4-8} 
\cline{10-14}
\colhead{Species} &
\colhead{$Z$} &
\colhead{$\log\varepsilon_{\odot}$} &
\colhead{$\log\varepsilon$(X)} &
\colhead{[X/Fe]} &
\colhead{$\sigma$($\log\varepsilon$(X))} &
\colhead{$\sigma$([X/Fe])} &
\colhead{$N$} &
\colhead{} &
\colhead{$\log\varepsilon$(X)} &
\colhead{[X/Fe]} &
\colhead{$\sigma$($\log\varepsilon$(X))} &
\colhead{$\sigma$([X/Fe])} &
\colhead{$N$} 
}
\startdata
Li~\textsc{i}  & 3 &\nodata& $<$0.25 & \nodata &\nodata&\nodata&  1 && $<$0.55 &\nodata&\nodata&\nodata&  1  \\
C (CH)         &  6 & 8.43 &    7.41 & $+$1.95 & 0.20  & 0.20  &  1 &&    5.45 & $+$0.36 & 0.20 & 0.20 &  1  \\
N (CN)         &  7 & 7.83 &    6.70 & $+$1.84 & 0.30  & 0.30  &  1 && \nodata &\nodata&\nodata&\nodata&  0  \\
O~\textsc{i}   &  8 & 8.69 &    8.08 & $+$2.36 & 0.15  & 0.18  &  2 && $<$6.70 &$<+$1.35&\nodata&\nodata& 1  \\
Na~\textsc{i}  & 11 & 6.24 &    4.86 & $+$1.59 & 0.13  & 0.14  &  4 &&    2.61 & $-$0.29 & 0.29 & 0.11 &  2  \\
Mg~\textsc{i}  & 12 & 7.60 &    6.47 & $+$1.84 & 0.21  & 0.13  &  3 &&    4.47 & $+$0.21 & 0.14 & 0.17 &  2  \\
Al~\textsc{i}  & 13 & 6.45 & $<$5.20 &$<+$1.72 &\nodata&\nodata&  2 &&    3.07 & $-$0.04 & 0.39 & 0.28 &  1  \\
Si~\textsc{i}  & 14 & 7.51 &    6.28 & $+$1.74 & 0.10  & 0.18  & 13 &&    4.27 & $+$0.10 & 0.38 & 0.36 &  1  \\
K~\textsc{i}   & 19 & 5.03 &    3.34 & $+$1.28 & 0.26  & 0.13  &  1 &&    2.13 & $+$0.44 & 0.22 & 0.13 &  2  \\
Ca~\textsc{i}  & 20 & 6.34 &    3.87 & $+$0.50 & 0.16  & 0.09  & 14 &&    3.15 & $+$0.15 & 0.18 & 0.12 &  6  \\
Sc~\textsc{ii} & 21 & 3.15 & $-$0.23 & $-$0.41 & 0.13  & 0.16  &  4 && $-$0.30 & $-$0.11 & 0.14 & 0.14 &  5  \\
Ti~\textsc{i}  & 22 & 4.95 &    2.40 & $+$0.42 & 0.27  & 0.07  & 11 &&    1.58 & $-$0.03 & 0.31 & 0.13 &  5  \\
Ti~\textsc{ii} & 22 & 4.95 &    1.81 & $-$0.17 & 0.11  & 0.14  &  9 &&    1.80 & $+$0.19 & 0.11 & 0.10 & 16  \\
V~\textsc{i}   & 23 & 3.93 & \nodata & \nodata &\nodata&\nodata&  0 && \nodata &\nodata&\nodata&\nodata&  0  \\
V~\textsc{ii}  & 23 & 3.93 & \nodata & \nodata &\nodata&\nodata&  0 && \nodata &\nodata&\nodata&\nodata&  0  \\
Cr~\textsc{i}  & 24 & 5.64 &    2.64 & $-$0.04 & 0.25  & 0.06  &  9 &&    2.01 & $-$0.29 & 0.27 & 0.12 &  4  \\
Cr~\textsc{ii} & 24 & 5.64 & \nodata & \nodata &\nodata&\nodata&  0 &&    2.40 & $+$0.10 & 0.33 & 0.33 &  1  \\
Mn~\textsc{i}  & 25 & 5.43 &    2.46 & $+$0.00 & 0.22  & 0.09  &  3 &&    1.77 & $-$0.32 & 0.28 & 0.20 &  1  \\
Fe~\textsc{i}  & 26 & 7.50 &    4.53 & $-$2.97 & 0.22  & 0.22  & 70 &&    4.16 & $-$3.34 & 0.26 & 0.26 & 78  \\
Fe~\textsc{ii} & 26 & 7.50 &    4.07 & $-$3.43 & 0.14  & 0.14  &  2 &&    4.47 & $-$3.03 & 0.11 & 0.11 &  7  \\
Co~\textsc{i}  & 27 & 4.99 & \nodata & \nodata &\nodata&\nodata&  0 && \nodata &\nodata&\nodata&\nodata&  0  \\
Ni~\textsc{i}  & 28 & 6.22 &    3.40 & $+$0.15 & 0.21  & 0.09  &  6 &&    2.50 & $-$0.38 & 0.25 & 0.10 &  1  \\
Zn~\textsc{i}  & 30 & 4.56 &    2.26 & $+$0.67 & 0.10  & 0.23  &  2 &&    1.67 & $+$0.45 & 0.24 & 0.34 &  2  \\
Sr~\textsc{ii} & 38 & 2.87 & $-$2.47 & $-$2.37 & 0.21  & 0.25  &  1 && $-$0.97 & $-$0.49 & 0.25 & 0.25 &  2  \\
Y~\textsc{ii}  & 39 & 2.21 & \nodata & \nodata &\nodata&\nodata&  0 && $-$1.34 & $-$0.21 & 0.19 & 0.19 &  3  \\
Zr~\textsc{ii} & 40 & 2.58 & \nodata & \nodata &\nodata&\nodata&  0 && $-$0.62 & $+$0.15 & 0.26 & 0.28 &  2  \\
Ba~\textsc{ii} & 56 & 2.18 & $-$2.24 & $-$1.45 & 0.17  & 0.20  &  3 && $-$1.27 & $-$0.10 & 0.16 & 0.16 &  4  \\
La~\textsc{ii} & 57 & 1.10 & \nodata & \nodata &\nodata&\nodata&  0 && $-$1.76 & $+$0.49 & 0.22 & 0.22 &  2  \\
Ce~\textsc{ii} & 58 & 1.58 & \nodata & \nodata &\nodata&\nodata&  0 && $-$1.35 & $+$0.42 & 0.50 & 0.50 &  2  \\
Pr~\textsc{ii} & 59 & 0.72 & \nodata & \nodata &\nodata&\nodata&  0 && \nodata &\nodata&\nodata&\nodata&  0  \\
Nd~\textsc{ii} & 60 & 1.42 & \nodata & \nodata &\nodata&\nodata&  0 && $-$1.41 & $+$0.51 & 0.25 & 0.26 &  2  \\
Sm~\textsc{ii} & 62 & 0.96 & \nodata & \nodata &\nodata&\nodata&  0 && $-$1.53 & $+$0.86 & 0.50 & 0.50 &  2  \\
Eu~\textsc{ii} & 63 & 0.52 &$<-$2.40 &$<+$0.05 &\nodata&\nodata&  2 && $-$2.12 & $+$0.70 & 0.20 & 0.21 &  3  \\
Dy~\textsc{ii} & 66 & 1.10 & \nodata & \nodata &\nodata&\nodata&  0 && \nodata &\nodata&\nodata&\nodata&  0  \\
Pb~\textsc{i}  & 82 & 2.04 & $<$0.60 &$<+$1.53 &\nodata&\nodata&  1 && $<$0.29 &$<+$1.59 &\nodata&\nodata&1  
\enddata      
\tablecomments{%
[Fe/H] is given instead of [X/Fe] for Fe.
The C abundances have been corrected 
(by $+$0.41 and $+$0.75~dex) to the ``natal'' abundances
according to the stellar evolution corrections presented by
\citet{placco14c}.
A single C abundance is derived by spectrum synthesis of the
region from 4290--4330~\AA.~
NLTE corrections have been applied to the 
Li, Na, Mg, Al, Si, K, Fe~\textsc{i}, and Pb abundances;
see Table~\ref{tab:lineabund} for corrections and
the text for references.
}
%\tablenotetext{a}{%
%}
\end{deluxetable*}

\begin{deluxetable*}{cccccccccccccc}
\tablecaption{Derived Abundances (Part 2 of 3)
\label{tab:abund2}}
\tablewidth{0pt}
\tabletypesize{\scriptsize}
\tablehead{
\colhead{} &
\colhead{} &
\colhead{} &
\multicolumn{5}{c}{J1015$-$0238} &
\colhead{} &
\multicolumn{5}{c}{J1018$-$0155} \\
\cline{4-8} 
\cline{10-14}
\colhead{Species} &
\colhead{$Z$} &
\colhead{$\log\varepsilon_{\odot}$} &
\colhead{$\log\varepsilon$(X)} &
\colhead{[X/Fe]} &
\colhead{$\sigma$($\log\varepsilon$(X))} &
\colhead{$\sigma$([X/Fe])} &
\colhead{$N$} &
\colhead{} &
\colhead{$\log\varepsilon$(X)} &
\colhead{[X/Fe]} &
\colhead{$\sigma$($\log\varepsilon$(X))} &
\colhead{$\sigma$([X/Fe])} &
\colhead{$N$} 
}
\startdata
Li~\textsc{i}  & 3 &\nodata& $<$0.21 &\nodata&\nodata&\nodata&   1 & & $<$0.51 &\nodata&\nodata&\nodata&   1 \\
C (CH)         &  6 & 8.43 &    5.77 & $-$0.22 & 0.20 & 0.20 &   1 & &    5.37 & $-$0.25 & 0.20 & 0.20 &   1 \\
N (CN)         &  7 & 7.83 & \nodata &\nodata&\nodata&\nodata&   0 & & \nodata &\nodata&\nodata &\nodata&  0 \\
O~\textsc{i}   &  8 & 8.69 & $<$6.70 &$<+$0.65&\nodata&\nodata&  1 & & $<$6.80 &$<+$0.92&\nodata&\nodata&  1 \\
Na~\textsc{i}  & 11 & 6.24 &    3.62 & $+$0.02 & 0.38 & 0.16 &   2 & &    3.50 & $+$0.07 & 0.35 & 0.15 &   2 \\
Mg~\textsc{i}  & 12 & 7.60 &    5.19 & $+$0.23 & 0.18 & 0.17 &   3 & &    5.11 & $+$0.32 & 0.14 & 0.14 &   2 \\
Al~\textsc{i}  & 13 & 6.45 &    3.92 & $+$0.11 & 0.39 & 0.25 &   1 & &    4.38 & $+$0.74 & 0.38 & 0.24 &   1 \\
Si~\textsc{i}  & 14 & 7.51 &    5.07 & $+$0.20 & 0.43 & 0.37 &   1 & &    4.71 & $+$0.01 & 0.40 & 0.36 &   1 \\
K~\textsc{i}   & 19 & 5.03 &    2.42 & $+$0.03 & 0.21 & 0.07 &   2 & &    2.40 & $+$0.18 & 0.20 & 0.07 &   2 \\
Ca~\textsc{i}  & 20 & 6.34 &    3.82 & $+$0.12 & 0.17 & 0.09 &  15 & &    3.65 & $+$0.12 & 0.16 & 0.10 &  12 \\
Sc~\textsc{ii} & 21 & 3.15 &    0.31 & $-$0.20 & 0.11 & 0.11 &   9 & &    0.17 & $-$0.17 & 0.11 & 0.10 &   9 \\
Ti~\textsc{i}  & 22 & 4.95 &    2.35 & $+$0.04 & 0.29 & 0.07 &  16 & &    2.08 & $-$0.06 & 0.29 & 0.09 &  10 \\
Ti~\textsc{ii} & 22 & 4.95 &    2.58 & $+$0.27 & 0.11 & 0.09 &  25 & &    2.27 & $+$0.13 & 0.09 & 0.09 &  20 \\
V~\textsc{i}   & 23 & 3.93 &    0.94 & $-$0.35 & 0.31 & 0.11 &   3 & &    0.87 & $-$0.25 & 0.32 & 0.16 &   3 \\
V~\textsc{ii}  & 23 & 3.93 &    1.22 & $-$0.07 & 0.17 & 0.17 &   2 & &    1.05 & $-$0.07 & 0.30 & 0.29 &   1 \\
Cr~\textsc{i}  & 24 & 5.64 &    2.77 & $-$0.23 & 0.27 & 0.07 &   7 & &    2.45 & $-$0.38 & 0.25 & 0.06 &   8 \\
Cr~\textsc{ii} & 24 & 5.64 &    3.07 & $+$0.07 & 0.15 & 0.12 &   2 & &    2.67 & $-$0.16 & 0.28 & 0.27 &   1 \\
Mn~\textsc{i}  & 25 & 5.43 &    2.38 & $-$0.41 & 0.23 & 0.09 &   3 & &    2.08 & $-$0.54 & 0.22 & 0.10 &   3 \\
Fe~\textsc{i}  & 26 & 7.50 &    4.86 & $-$2.64 & 0.24 & 0.24 & 115 & &    4.69 & $-$2.81 & 0.24 & 0.24 & 107 \\
Fe~\textsc{ii} & 26 & 7.50 &    4.77 & $-$2.73 & 0.11 & 0.11 &  10 & &    4.61 & $-$2.89 & 0.11 & 0.11 &  11 \\
Co~\textsc{i}  & 27 & 4.99 &    2.06 & $-$0.29 & 0.34 & 0.16 &   1 & &    1.84 & $-$0.34 & 0.31 & 0.14 &   1 \\
Ni~\textsc{i}  & 28 & 6.22 &    3.56 & $-$0.02 & 0.21 & 0.08 &   6 & &    3.30 & $-$0.11 & 0.23 & 0.13 &   5 \\
Zn~\textsc{i}  & 30 & 4.56 &    2.24 & $+$0.32 & 0.11 & 0.24 &   2 & &    1.83 & $+$0.08 & 0.17 & 0.27 &   2 \\
Sr~\textsc{ii} & 38 & 2.87 &    0.31 & $+$0.08 & 0.18 & 0.19 &   2 & & $-$0.72 & $-$0.78 & 0.25 & 0.25 &   2 \\
Y~\textsc{ii}  & 39 & 2.21 & $-$0.80 & $-$0.37 & 0.12 & 0.13 &   3 & & $-$1.39 & $-$0.79 & 0.16 & 0.17 &   2 \\
Zr~\textsc{ii} & 40 & 2.58 &    0.01 & $+$0.07 & 0.15 & 0.15 &   3 & & $-$0.52 & $-$0.29 & 0.21 & 0.21 &   3 \\
Ba~\textsc{ii} & 56 & 2.18 & $-$1.59 & $-$1.13 & 0.15 & 0.16 &   4 & & $-$0.94 & $-$0.31 & 0.15 & 0.15 &   4 \\
La~\textsc{ii} & 57 & 1.10 & \nodata &\nodata&\nodata&\nodata&   0 & & $-$1.71 & $+$0.00 & 0.23 & 0.24 &   5 \\
Ce~\textsc{ii} & 58 & 1.58 & \nodata &\nodata&\nodata&\nodata&   0 & & $-$1.20 & $+$0.03 & 0.26 & 0.27 &   2 \\
Pr~\textsc{ii} & 59 & 0.72 & \nodata &\nodata&\nodata&\nodata&   0 & & $-$1.62 & $+$0.47 & 0.27 & 0.28 &   1 \\
Nd~\textsc{ii} & 60 & 1.42 & \nodata &\nodata&\nodata&\nodata&   0 & & $-$1.15 & $+$0.24 & 0.20 & 0.20 &   3 \\
Sm~\textsc{ii} & 62 & 0.96 & \nodata &\nodata&\nodata&\nodata&   0 & & $-$1.73 & $+$0.12 & 0.45 & 0.45 &   1 \\
Eu~\textsc{ii} & 63 & 0.52 &$<-$2.70 &$<-$0.58&\nodata&\nodata&  2 & & $-$1.96 & $+$0.33 & 0.15 & 0.15 &   3 \\
Dy~\textsc{ii} & 66 & 1.10 & \nodata &\nodata&\nodata&\nodata&   0 & & $-$1.06 & $+$0.65 & 0.43 & 0.44 &   1 \\
Pb~\textsc{i}  & 82 & 2.04 & $<$0.10 &$<+$0.70&\nodata&\nodata&  1 & & $<$0.05 &$<+$0.82&\nodata&\nodata&  1 
\enddata      
\tablecomments{%
[Fe/H] is given instead of [X/Fe] for Fe.
The C abundances have been corrected 
(by $+$0.77 and $+$0.76~dex) to the ``natal'' abundances
according to the stellar evolution corrections presented by
\citet{placco14c}.
A single C abundance is derived by spectrum synthesis of the
region from 4290--4330~\AA.~
NLTE corrections have been applied to the 
Li, Na, Mg, Al, Si, K, Fe~\textsc{i}, and Pb abundances;
see Table~\ref{tab:lineabund} for corrections and
the text for references.
}
%\tablenotetext{a}{%
%}
\end{deluxetable*}

\begin{deluxetable*}{cccccccc}
\tablecaption{Derived Abundances (Part 3 of 3)
\label{tab:abund3}}
\tablewidth{0pt}
\tabletypesize{\scriptsize}
\tablehead{
\colhead{} &
\colhead{} &
\colhead{} &
\multicolumn{5}{c}{J1018$-$0209} \\
\cline{4-8} 
\colhead{Species} &
\colhead{$Z$} &
\colhead{$\log\varepsilon_{\odot}$} &
\colhead{$\log\varepsilon$(X)} &
\colhead{[X/Fe]} &
\colhead{$\sigma$($\log\varepsilon$(X))} &
\colhead{$\sigma$([X/Fe])} &
\colhead{$N$} 
}
\startdata
Li~\textsc{i}  & 3 &\nodata& $<$0.25 & \nodata &\nodata&\nodata&   1 \\
C (CH)         &  6 & 8.43 &    5.34 & $-$0.34 & 0.20  & 0.20  &   1 \\
N (CN)         &  7 & 7.83 & \nodata & \nodata &\nodata&\nodata&   0 \\
O~\textsc{i}   &  8 & 8.69 & $<$6.90 &$<+$0.96 &\nodata&\nodata&   1 \\
Na~\textsc{i}  & 11 & 6.24 &    3.25 & $-$0.24 & 0.24  & 0.08  &   3 \\
Mg~\textsc{i}  & 12 & 7.60 &    5.15 & $+$0.30 & 0.15  & 0.13  &   5 \\
Al~\textsc{i}  & 13 & 6.45 &    3.30 & $-$0.40 & 0.40  & 0.28  &   1 \\
Si~\textsc{i}  & 14 & 7.51 &    4.98 & $+$0.22 & 0.26  & 0.26  &   2 \\
K~\textsc{i}   & 19 & 5.03 &    2.36 & $+$0.08 & 0.21  & 0.09  &   1 \\
Ca~\textsc{i}  & 20 & 6.34 &    3.63 & $+$0.04 & 0.16  & 0.10  &  13 \\
Sc~\textsc{ii} & 21 & 3.15 &    0.08 & $-$0.32 & 0.11  & 0.11  &   8 \\
Ti~\textsc{i}  & 22 & 4.95 &    2.05 & $-$0.15 & 0.30  & 0.07  &  11 \\
Ti~\textsc{ii} & 22 & 4.95 &    2.19 & $-$0.01 & 0.10  & 0.09  &  20 \\
V~\textsc{i}   & 23 & 3.93 &    0.60 & $-$0.58 & 0.34  & 0.19  &   1 \\
V~\textsc{ii}  & 23 & 3.93 &    1.18 & $+$0.00 & 0.23  & 0.23  &   2 \\
Cr~\textsc{i}  & 24 & 5.64 &    2.53 & $-$0.36 & 0.27  & 0.06  &   5 \\
Cr~\textsc{ii} & 24 & 5.64 & \nodata & \nodata &\nodata&\nodata&   0 \\
Mn~\textsc{i}  & 25 & 5.43 &    2.07 & $-$0.61 & 0.22  & 0.09  &   3 \\
Fe~\textsc{i}  & 26 & 7.50 &    4.75 & $-$2.75 & 0.25  & 0.25  & 113 \\
Fe~\textsc{ii} & 26 & 7.50 &    4.64 & $-$2.86 & 0.10  & 0.10  &  13 \\
Co~\textsc{i}  & 27 & 4.99 & \nodata & \nodata &\nodata&\nodata&   0 \\
Ni~\textsc{i}  & 28 & 6.22 &    3.32 & $-$0.15 & 0.23  & 0.07  &   4 \\
Zn~\textsc{i}  & 30 & 4.56 &    2.00 & $+$0.19 & 0.12  & 0.26  &   2 \\
Sr~\textsc{ii} & 38 & 2.87 & $-$0.73 & $-$0.85 & 0.25  & 0.23  &   2 \\
Y~\textsc{ii}  & 39 & 2.21 & $-$1.21 & $-$0.67 & 0.14  & 0.15  &   3 \\
Zr~\textsc{ii} & 40 & 2.58 & $-$0.58 & $-$0.41 & 0.23  & 0.24  &   1 \\
Ba~\textsc{ii} & 56 & 2.18 & $-$0.87 & $-$0.30 & 0.18  & 0.17  &   5 \\
La~\textsc{ii} & 57 & 1.10 & $-$1.77 & $-$0.12 & 0.19  & 0.20  &   2 \\
Ce~\textsc{ii} & 58 & 1.58 & $-$1.37 & $-$0.20 & 0.50  & 0.50  &   2 \\
Pr~\textsc{ii} & 59 & 0.72 & \nodata & \nodata &\nodata&\nodata&   0 \\
Nd~\textsc{ii} & 60 & 1.42 & \nodata & \nodata &\nodata&\nodata&   0 \\
Sm~\textsc{ii} & 62 & 0.96 & $-$1.64 & $+$0.15 & 0.50  & 0.50  &   2 \\
Eu~\textsc{ii} & 63 & 0.52 & $-$2.06 & $+$0.17 & 0.20  & 0.21  &   3 \\
Dy~\textsc{ii} & 66 & 1.10 & \nodata & \nodata &\nodata&\nodata&   0 \\
Pb~\textsc{i}  & 82 & 2.04 & $<$0.00 &$<+$0.71 &\nodata&\nodata&   1 \\
\enddata      
\tablecomments{%
[Fe/H] is given instead of [X/Fe] for Fe.
The C abundance has been corrected 
(by $+$0.77~dex) to the ``natal'' abundance
according to the stellar evolution corrections presented by
\citet{placco14c}.
A single C abundance is derived by spectrum synthesis of the
region from 4290--4330~\AA.~
NLTE corrections have been applied to the 
Li, Na, Mg, Al, Si, K, Fe~\textsc{i}, and Pb abundances;
see Table~\ref{tab:lineabund} for corrections and
the text for references.
}
%\tablenotetext{a}{%
%}
\end{deluxetable*}

\subsection{Abundance Derivations}
\label{sec:abund}

We use the MOOG ``abfind'' driver to derive abundances from
EWs of
Mg~\textsc{i}, Ca~\textsc{i}, Ti~\textsc{i} and \textsc{ii},
Cr~\textsc{i} and \textsc{ii}, Fe~\textsc{i} and \textsc{ii},
Ni~\textsc{i}, and some Zn~\textsc{i} lines.
Lines of these species are unblended,
are comprised of a single
or dominant isotope or do not exhibit any
significant line broadening by isotope shifts (IS),
and do not exhibit any significant 
line broadening by hyperfine structure (HFS).
All other abundances are derived by matching synthetic spectra
generated using the MOOG ``synth'' driver 
to the observed spectrum.
We produce line lists for these synthetic spectra
using the LINEMAKE code \citep{placco21linemake}.
We assume $^{12}$C/$^{13}$C = 4,
all N is $^{14}$N, and
\rpro\ isotopic ratios \citep{sneden08} in our syntheses.
Upper limits (U.L.)\ are reported for a few key species
based on the non-detection of one or more lines
in our spectra.
Table~\ref{tab:lineabund} reports the wavelengths ($\lambda$),
excitation potentials (E.P.), \loggf\ values and their references,
along with the EWs and LTE abundances
for each line in each star.

We apply NLTE corrections, when available and potentially non-negligible,
to the LTE abundances of each line of
Li~\textsc{i} \citep{lind09},
Na~\textsc{i} \citep{lind11}, 
Mg~\textsc{i} \citep{osorio15,osorio16}, 
Al~\textsc{i} \citep{nordlander17al}, 
Si~\textsc{i} \citep{shi09},
K~\textsc{i} \citep{takeda02},
and
Pb~\textsc{i} \citep{mashonkina12}.
The Li~\textsc{i}, 
Na~\textsc{i}, and 
Mg~\textsc{i} NLTE corrections
are accessed through the INSPECT database.
The stellar parameters occasionally lie beyond the
edge of pre-computed grids
(usually in \teff\ or \logg, with edges at 4500~K or 1.0~dex, respectively),
and in these cases we adopt the correction at the 
nearest point on the grid.
Table~\ref{tab:lineabund} lists
the line-by-line NLTE corrections, 
and Tables~\ref{tab:abund1}--\ref{tab:abund3} 
list the NLTE-corrected mean abundances.

We compute abundance uncertainties by drawing $10^{3}$ resamples of the
model atmosphere parameters, \loggf\ values, and
EWs (or approximations to the 
EWs for lines whose abundance was derived using spectrum synthesis),
assuming Gaussian uncertainties.
The uncertainties on the model atmosphere parameters 
are discussed in Section~\ref{sec:modelatm}.
The uncertainties in the \loggf\ values are taken from the 
grades assigned by the
National Institutes of Standards and Technology (NIST)
Atomic Spectra Database (ASD, version 5.9; \citealt{kramida21})
or the original source references
listed in Table~\ref{tab:lineabund}.
We assume a 5\% uncertainty in the EWs,
or a 5~m\AA\ minimum uncertainty in the case of weak lines,
which accounts for continuum placement and
unidentified weak blends.
We also include a wavelength-dependent component 
of EW uncertainty that reflects the
low S/N at blue wavelengths, which we empirically determine to be
$\sigma_{\rm EW} = 10^{24} \lambda^{-6.4}$,
where the wavelength, $\lambda$, is measured in \AA\ and 
$\sigma_{\rm EW}$ is measured in m\AA.~
This component of the uncertainty is 
$\approx 9$~m\AA\ at 4000~\AA,
$\approx 4$~m\AA\ at 4500~\AA,
$\approx 2$~m\AA\ at 5000~\AA, 
and
$<$~1 m\AA\ at 6000~\AA.
The mean abundance of each element is recomputed for each resample,
and the final abundance uncertainties reported in 
Tables~\ref{tab:abund1}--\ref{tab:abund3}
represent the 16th and 84th percentiles (i.e., 1$\sigma$ range)
of the distributions, which are roughly symmetric in most cases.

The uncertainties are generally smallest when the abundance
is derived from several lines with $\lambda \gtrsim$~4500~\AA,
where the S/N is highest.
There are several heavy elements,
including Ce, Pr, Sm, and Dy,
whose abundances are derived
from a small number (1 or 2) 
of very weak (EW $<$~10~m\AA\ or so) lines in the 
blue part of the spectrum ($\lambda <$~4500~\AA).~
The abundances are in agreement when
multiple lines of one of these elements are detected in a star,
which boosts our confidence in the legitimacy of their detection
despite the relatively large uncertainties.

\section{Results}
\label{sec:abundresults}

\begin{figure*}
\begin{center}
\includegraphics[angle=0,width=1.95in]{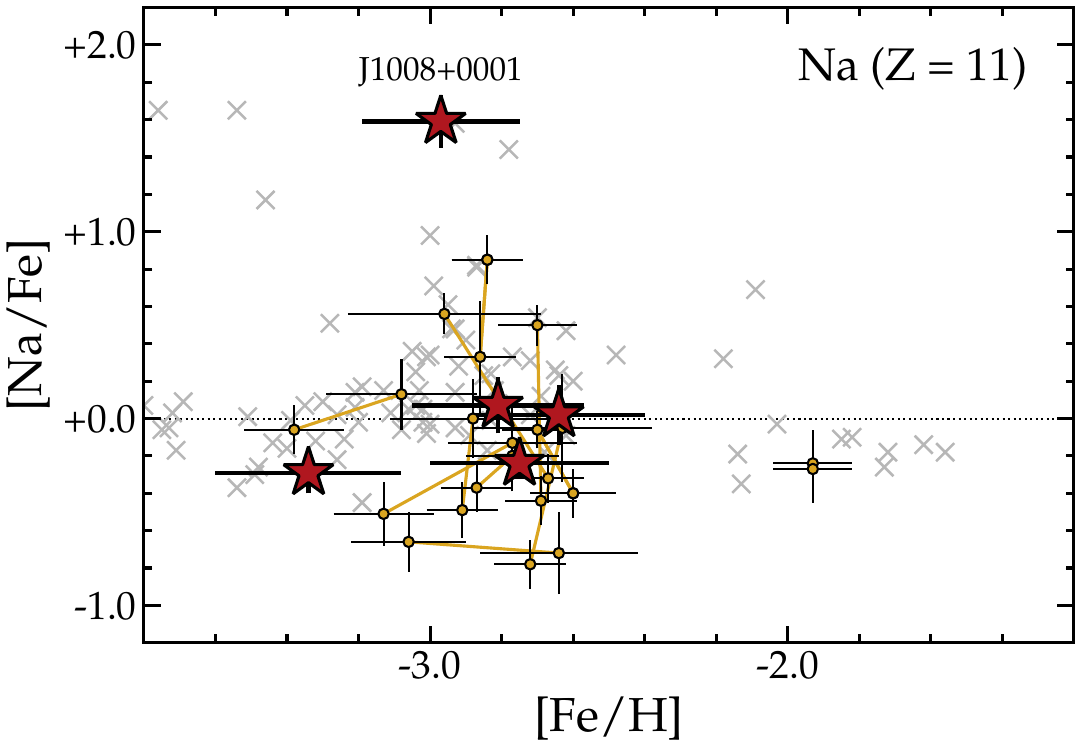}
\hspace*{0.1in}
\includegraphics[angle=0,width=1.95in]{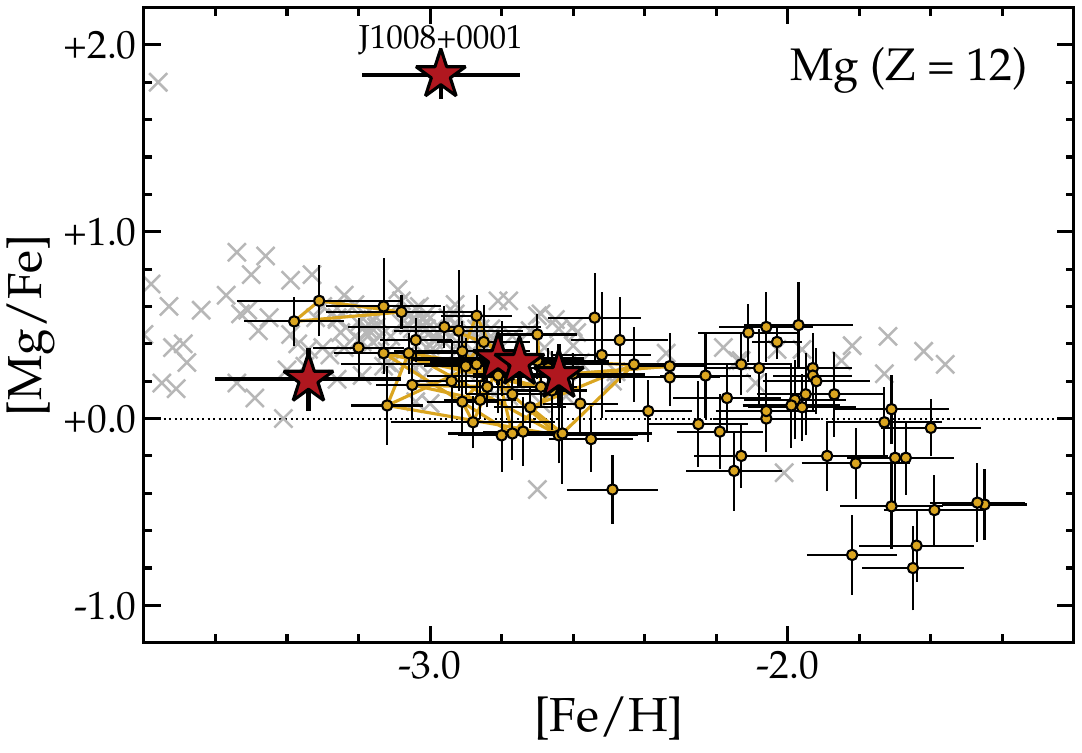}
\hspace*{0.1in}
\includegraphics[angle=0,width=1.95in]{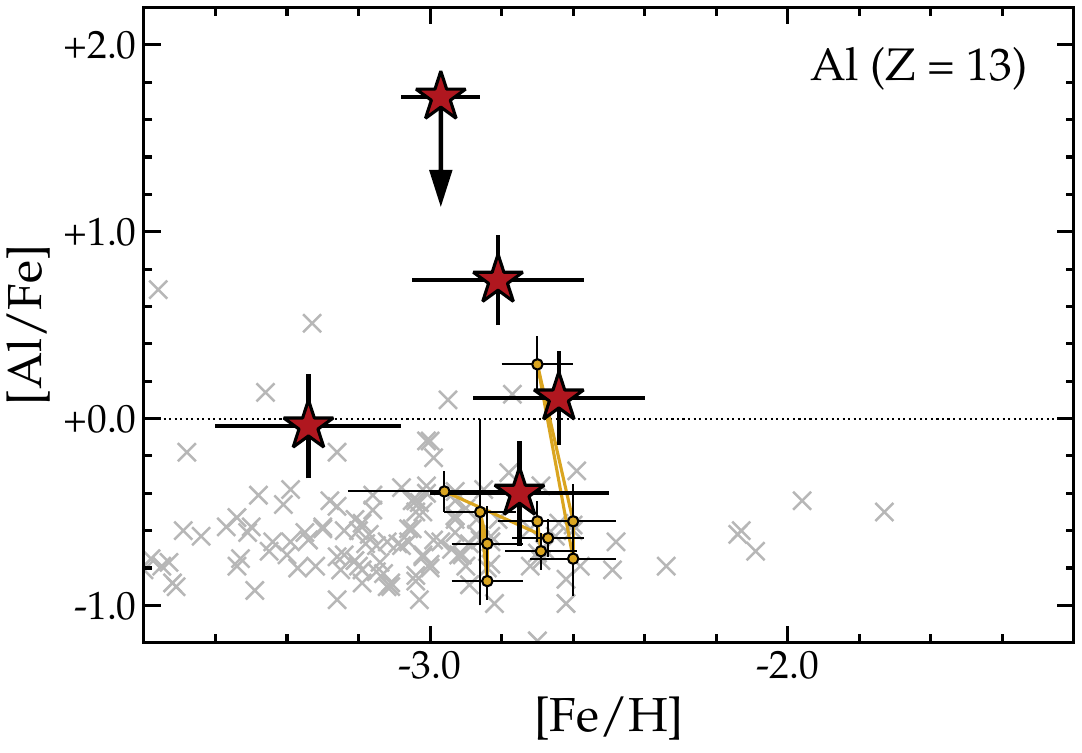} \\
\includegraphics[angle=0,width=1.95in]{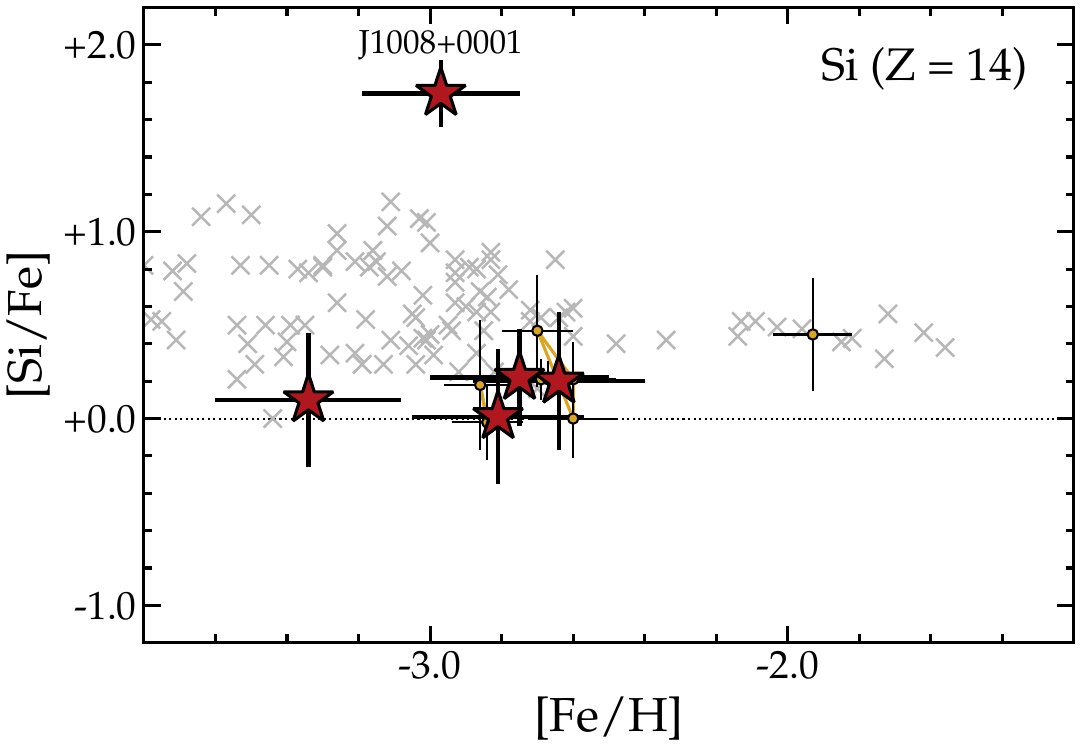}
\hspace*{0.1in}
\includegraphics[angle=0,width=1.95in]{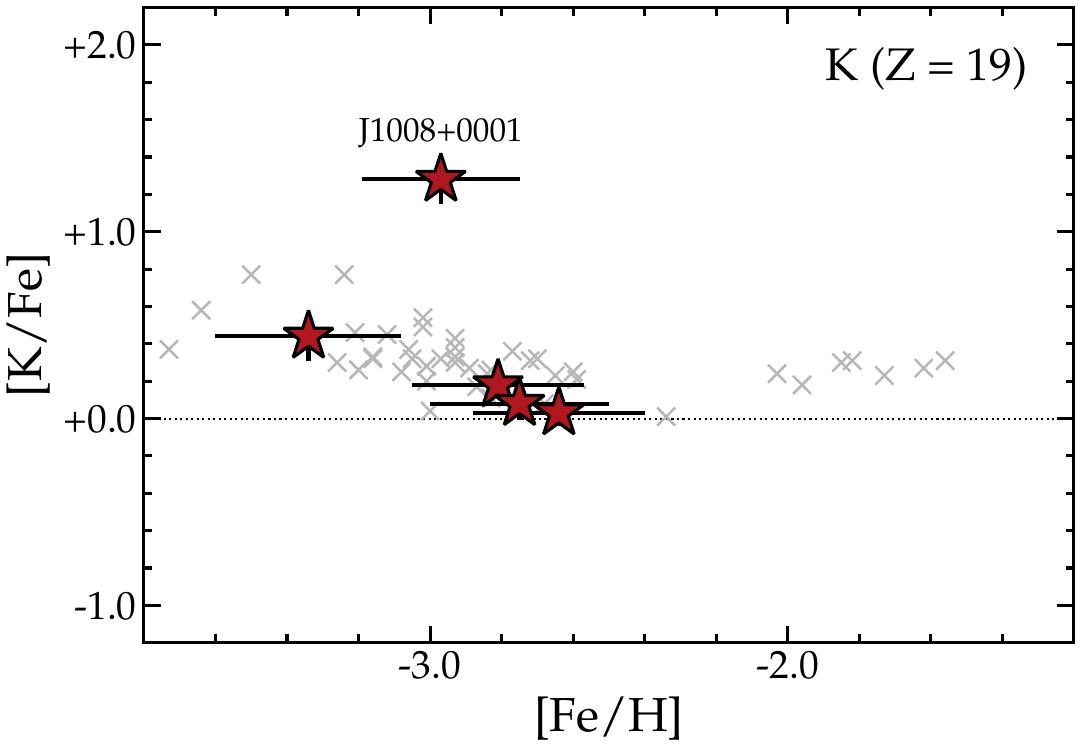}
\hspace*{0.1in}
\includegraphics[angle=0,width=1.95in]{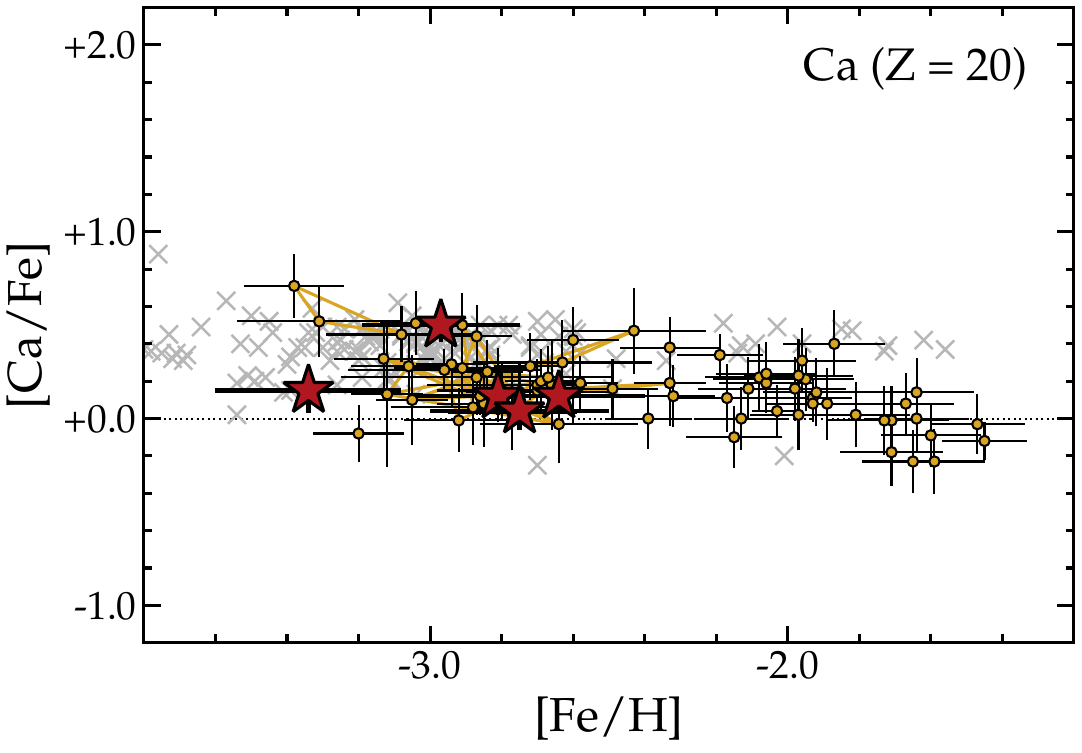} \\
\includegraphics[angle=0,width=1.95in]{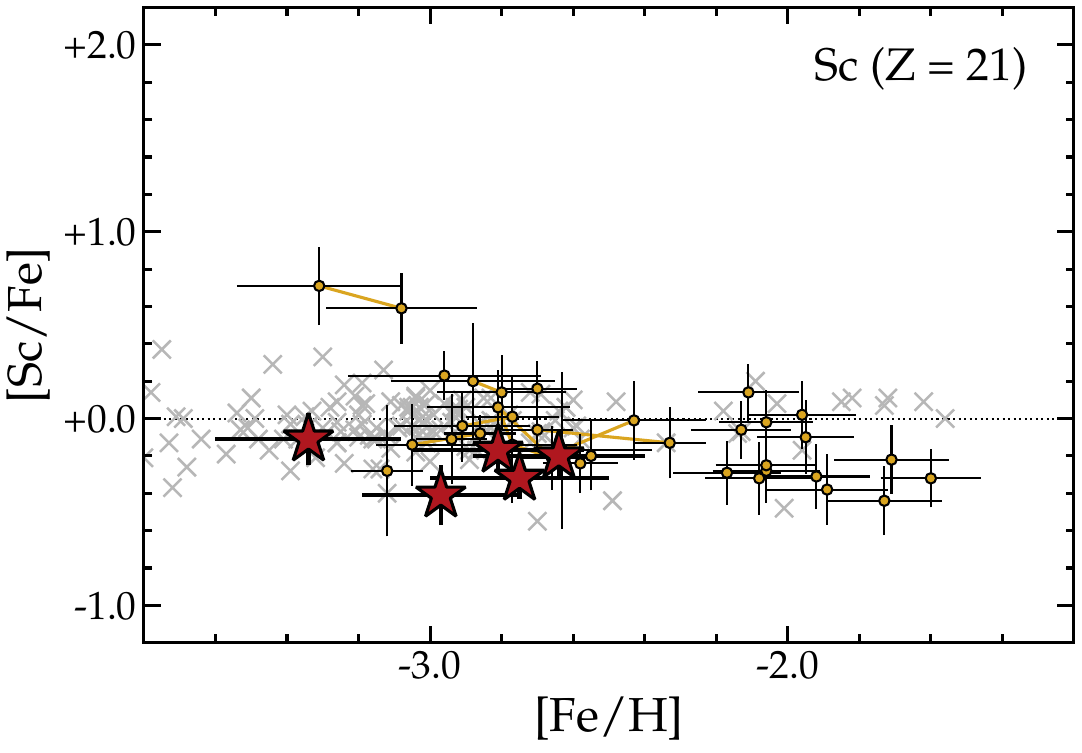}
\hspace*{0.1in}
\includegraphics[angle=0,width=1.95in]{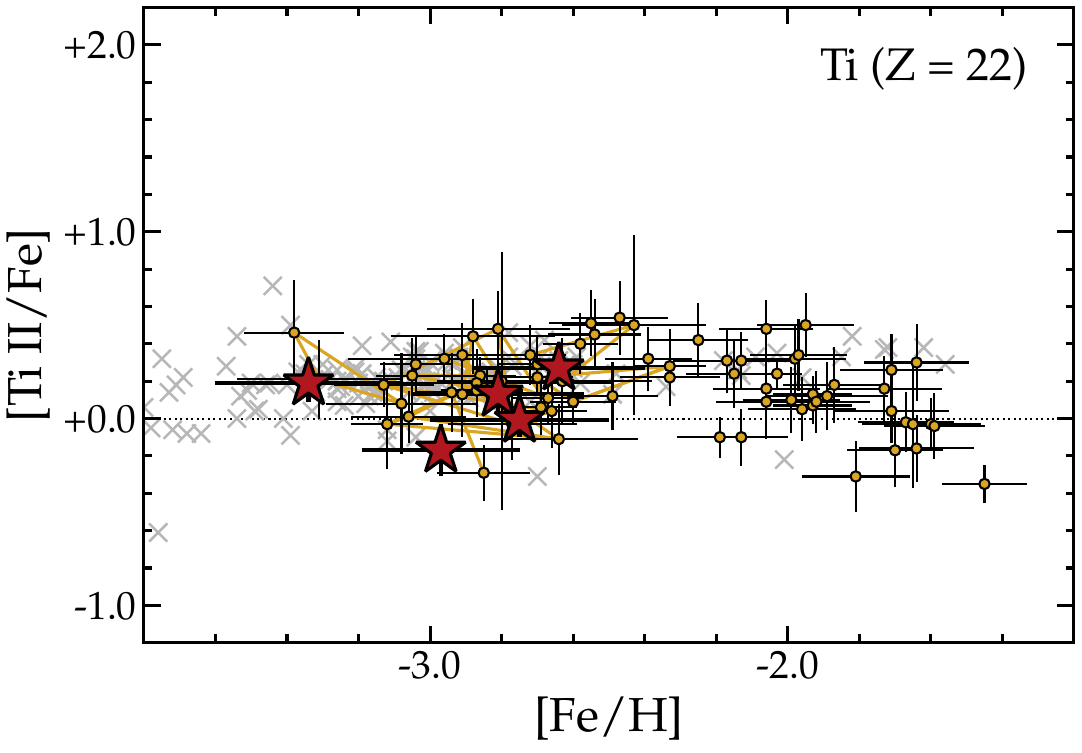}
\hspace*{0.1in}
\includegraphics[angle=0,width=1.95in]{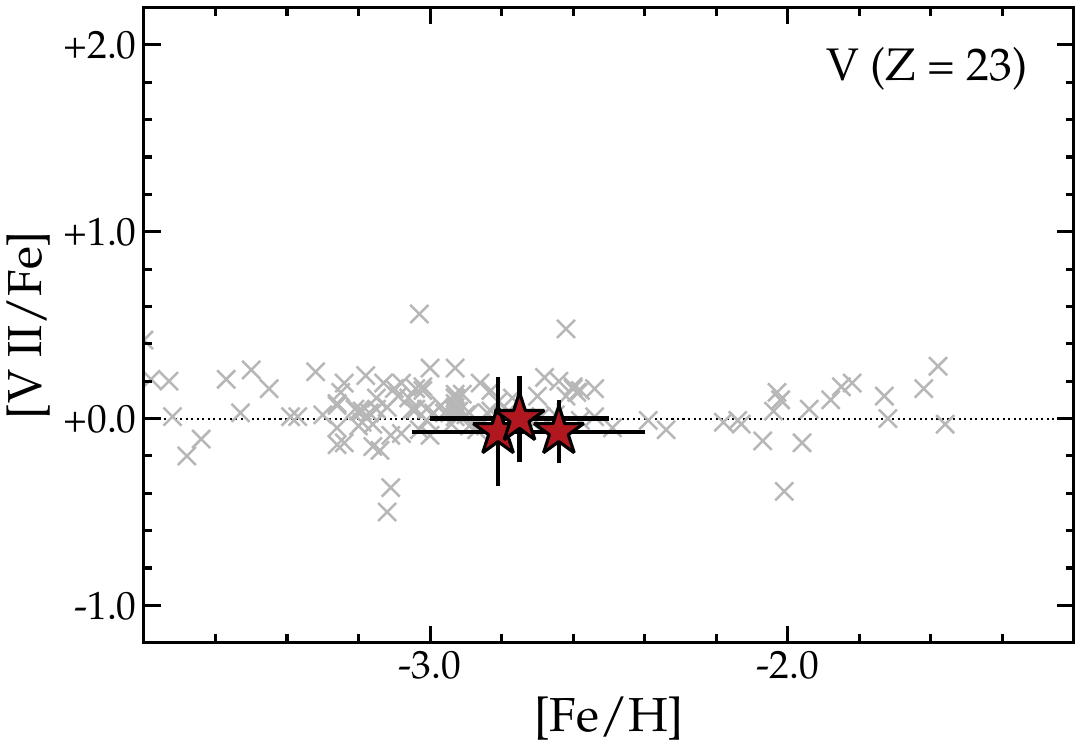} \\
\includegraphics[angle=0,width=1.95in]{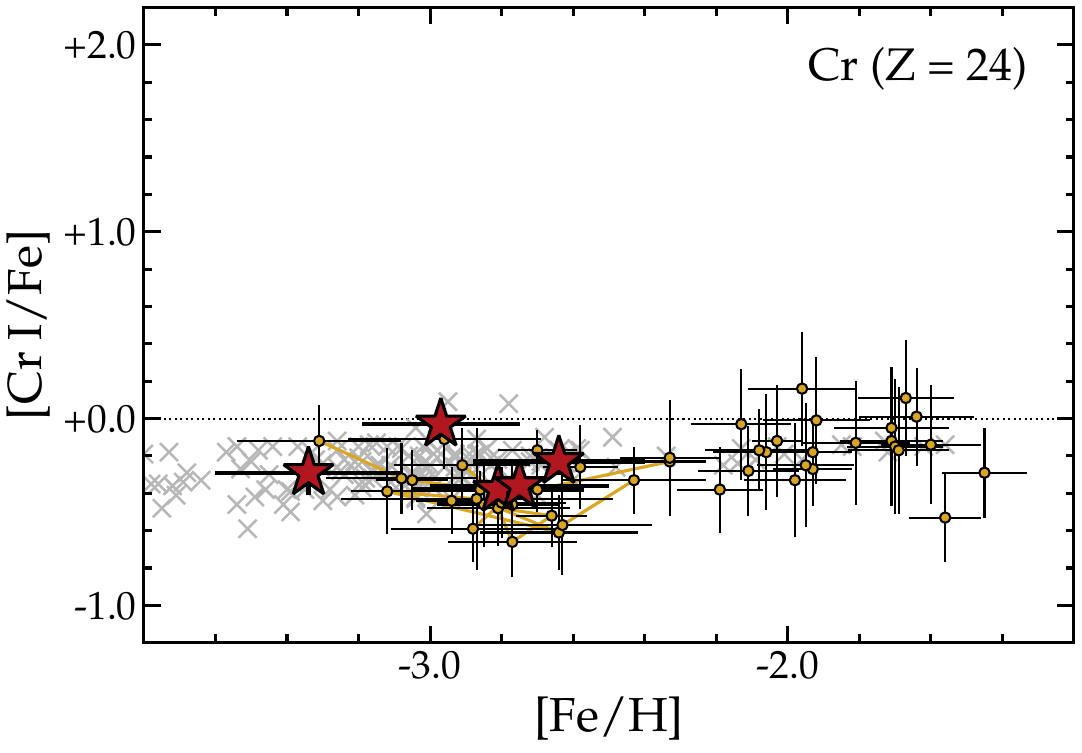}
\hspace*{0.1in}
\includegraphics[angle=0,width=1.95in]{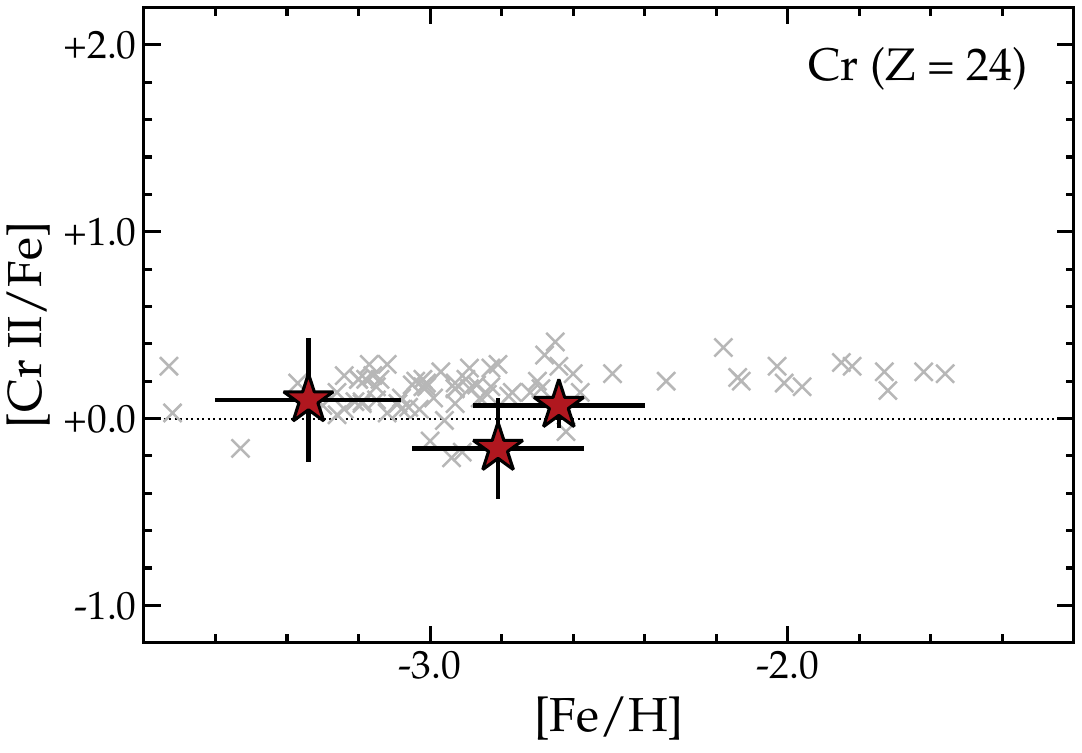}
\hspace*{0.1in}
\includegraphics[angle=0,width=1.95in]{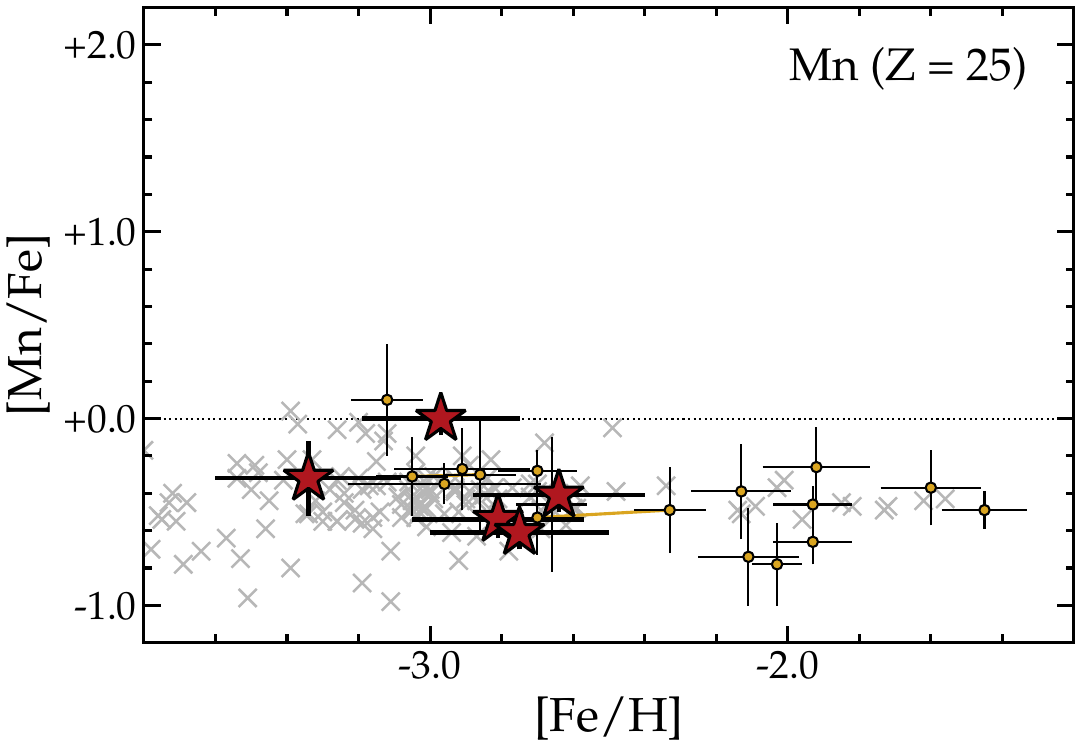} \\
\includegraphics[angle=0,width=1.95in]{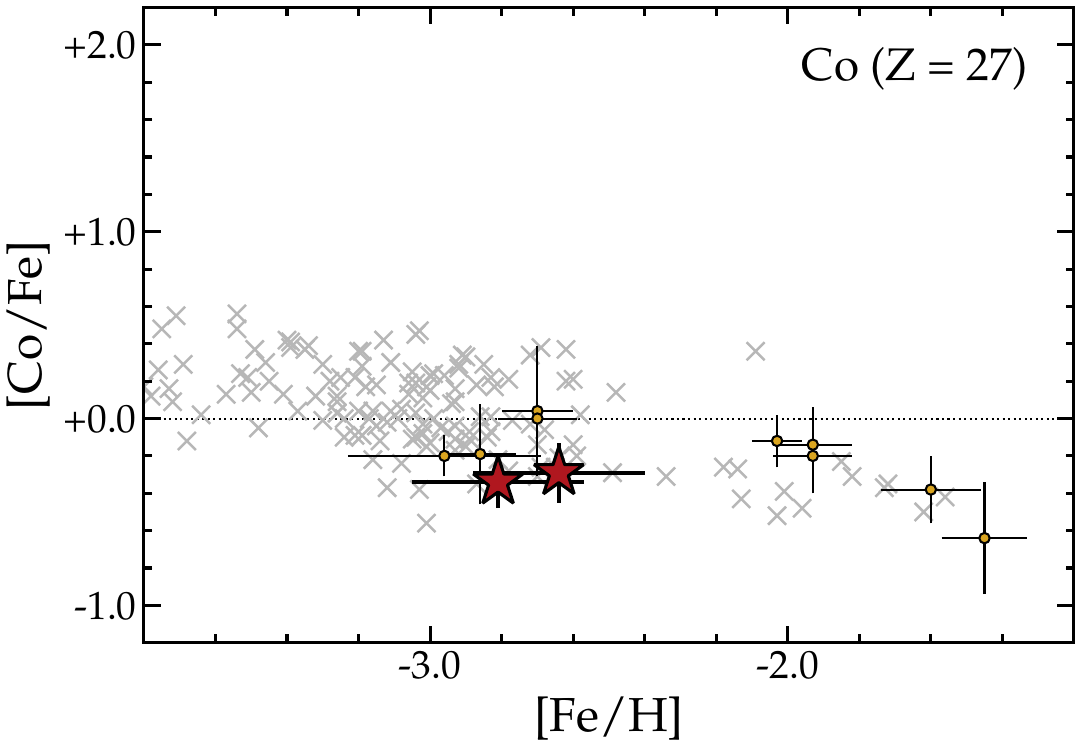}
\hspace*{0.1in}
\includegraphics[angle=0,width=1.95in]{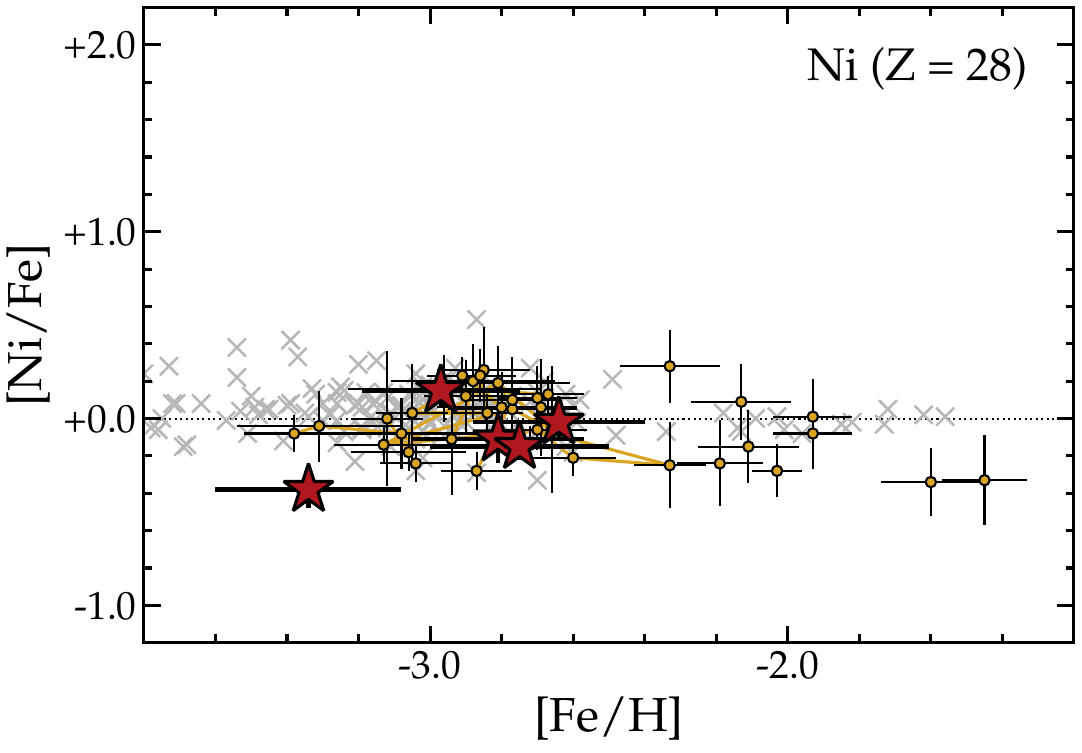}
\hspace*{0.1in}
\includegraphics[angle=0,width=1.95in]{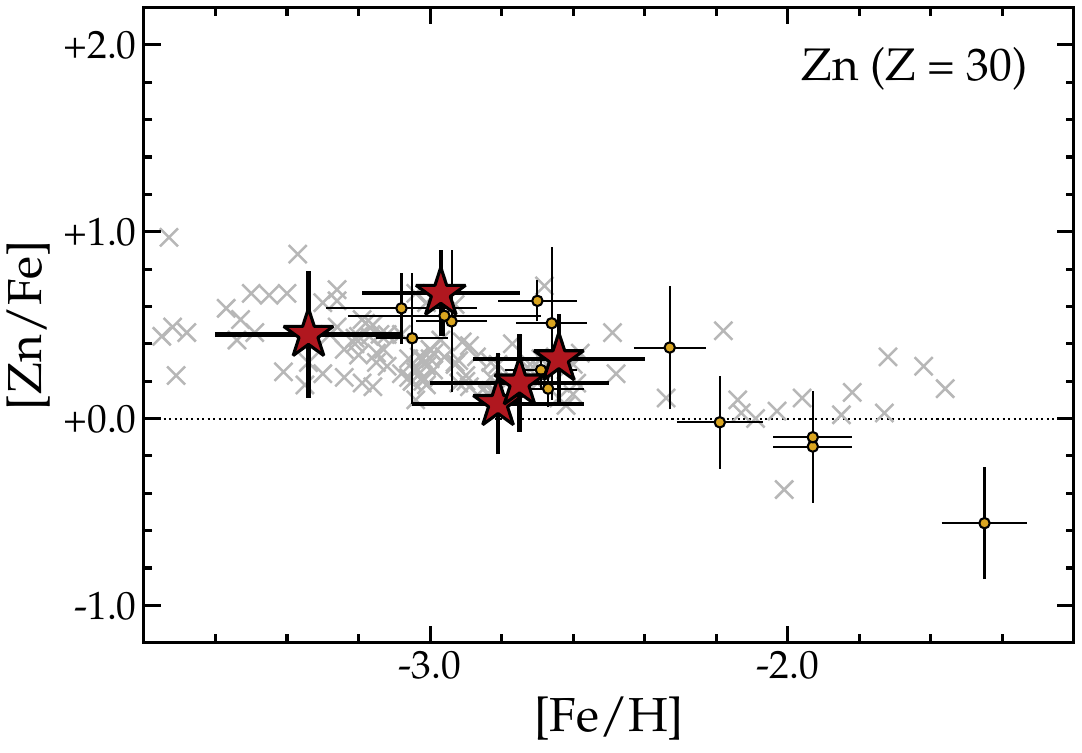} \\
\includegraphics[angle=0,width=1.95in]{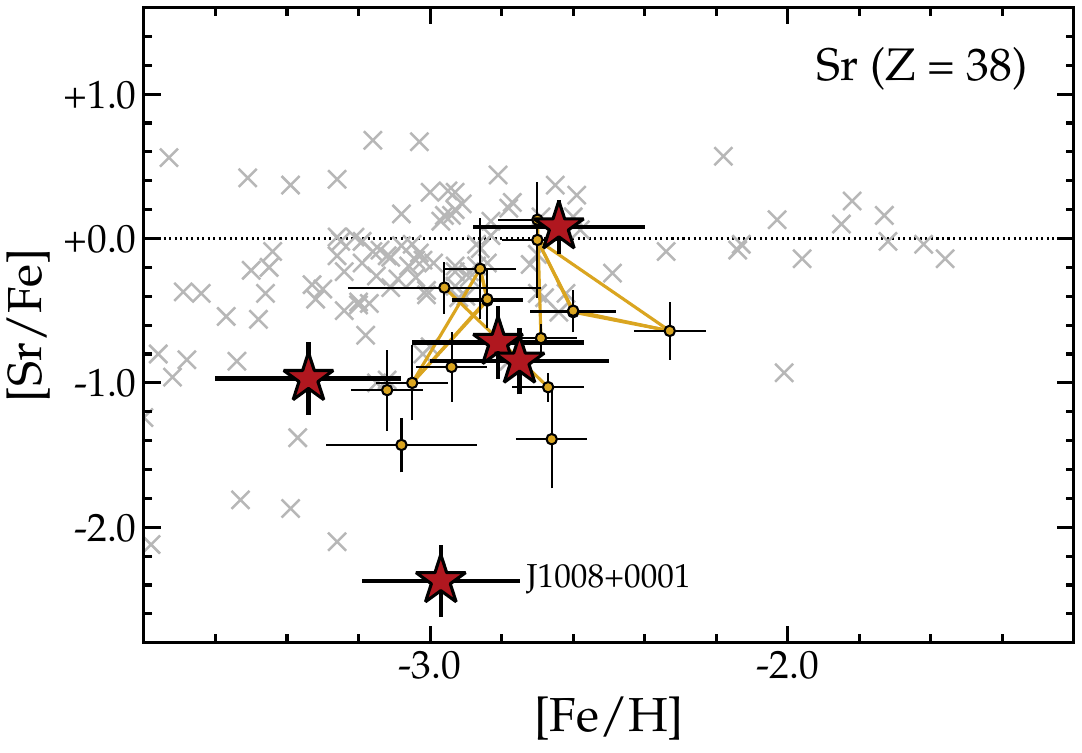}
\hspace*{0.1in}
\includegraphics[angle=0,width=1.95in]{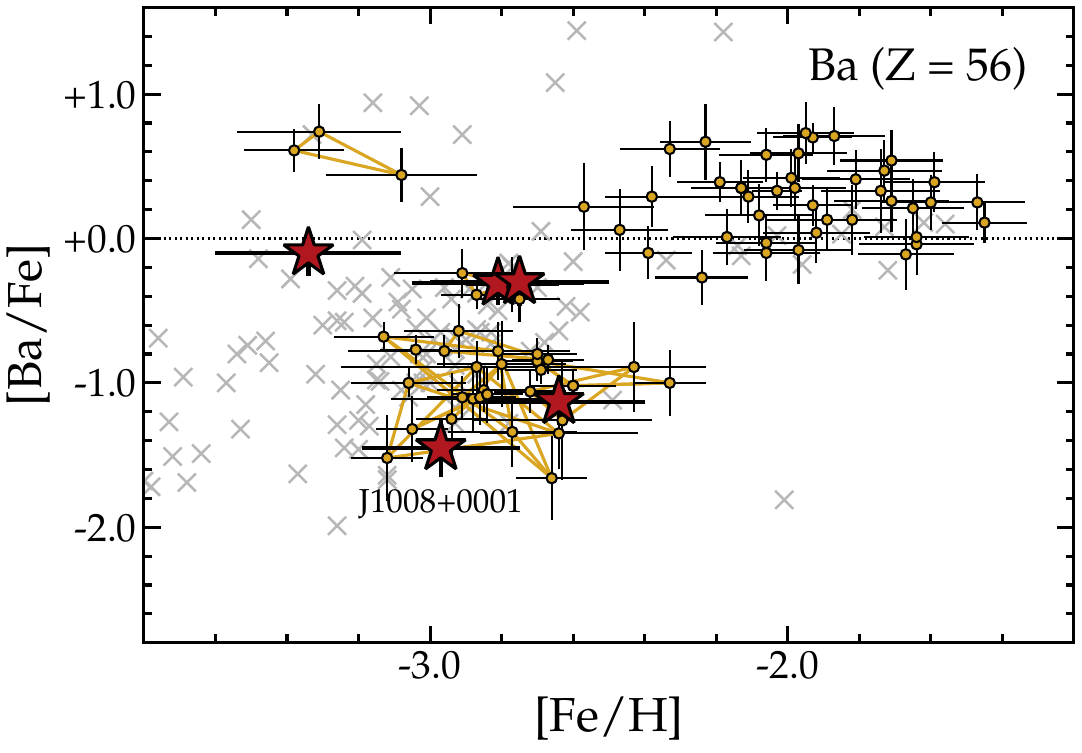}
\hspace*{0.1in} 
\includegraphics[angle=0,width=1.95in]{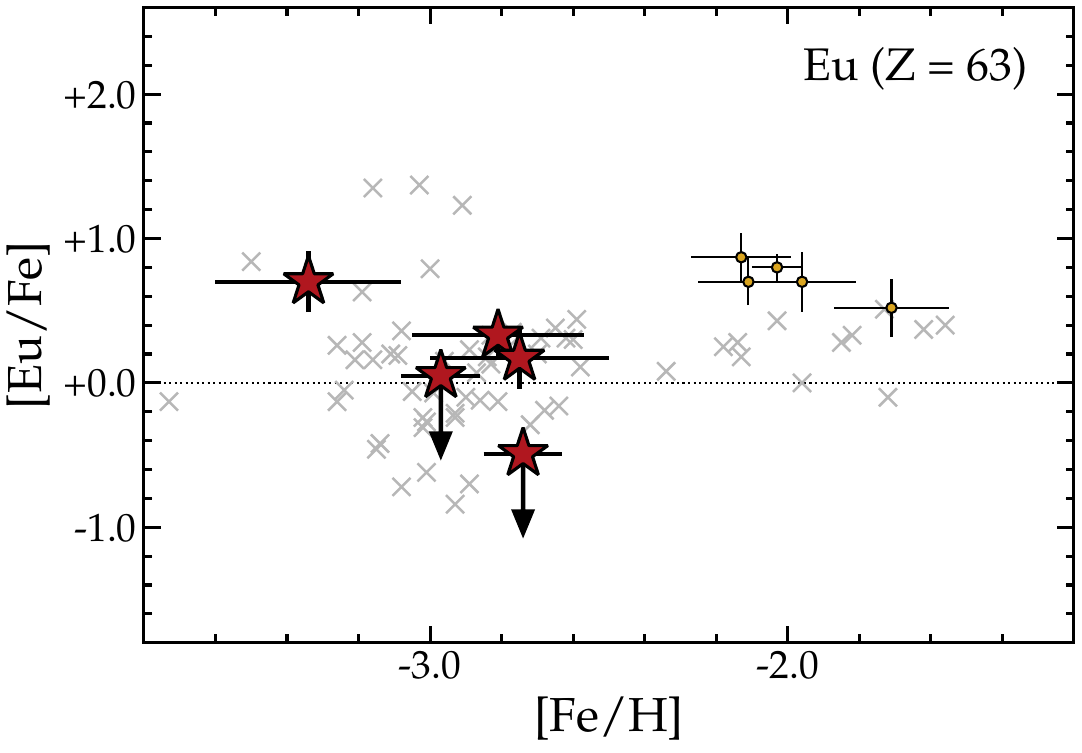}
\end{center}
\caption{
\label{fig:abundplot}
Abundance ratios for stars in our sample (large red stars)
compared with previous results for \sextans\ stars (yellow circles)
and metal-poor field stars (gray crosses).
Yellow lines connect stars reobserved or reanalyzed by previous studies.
The \sextans\ sample includes results from
\citet{shetrone01}, \citet{aoki09sex}, \citet{tafelmeyer10},
\citet{honda11sex}, \citet{mashonkina17dsph2,mashonkina22}, 
\citet{aoki20}, and \citet{lucchesi20}.
The field sample includes red giants
(\teff\ $<$ 5400~K) from 
\citet{cayrel04}, \citet{lai08}, \citet{yong13a}, 
\citet{roederer14c}, and \citet{ou20}.
Abundances in the comparison samples have been 
computed in LTE, except for Na in 
most studies, and
most abundances in the 11~stars 
studied by \citet{mashonkina17dsph2,mashonkina22}.
The panels are arranged in order of increasing atomic number ($Z$).
}
\end{figure*}

In this section we present our abundance results
and compare with previous work.
Our sample
contains no stars in common with previous 
high-resolution abundance studies.
Figure~\ref{fig:abundplot} shows the abundance ratios
for the stars in our sample,
previous results for \sextans\ stars, and 
metal-poor field stars in the solar neighborhood.
Several studies 
have reobserved or reanalyzed spectra of \sextans\ stars.
We display these results in Figure~\ref{fig:abundplot}
with lines connecting the different results for individual stars.

\subsection{$\alpha$ Elements:\ O, Mg, Si, Ca, and Ti}
\label{sec:alpha}

We detect five $\alpha$ elements in our sample:\
O ($Z = 8$), 
Mg ($Z = 12$), 
Si ($Z = 14$), 
Ca ($Z = 20$), and
Ti ($Z = 22$).
We detect O only in \sexcempno,
which we discuss separately in Section~\ref{sec:cempno}.
The mean [Mg/Fe], [Si/Fe], [Ca/Fe], and [Ti/Fe]
ratios found in the 
other four stars in our sample,
weighted by their inverse-squared uncertainties, are
$+0.27 \pm 0.08$, 
$+0.15 \pm 0.16$, 
$+0.10 \pm 0.05$, and
$+0.14 \pm 0.05$, respectively.
The weighted mean [$\alpha$/Fe] ratio in these four stars is
$+0.12 \pm 0.03$.
As shown in Figure~\ref{fig:abundplot},
these ratios are enhanced relative to the solar ratios,
but they are a few tenths of a dex low relative to the
mean ratios in field red giants with similar low metallicities.
These [$\alpha$/Fe] ratios
could indicate a deficiency of metals produced by the
highest-mass stars (e.g., \citealt{mcwilliam13}).

Our result is broadly consistent with abundances derived previously from
high-resolution spectra of the most metal-poor \sextans\ stars known
\citep{shetrone01,aoki09sex,tafelmeyer10,aoki20,mashonkina22}.
Our mean [Mg/Fe] abundance is in agreement with that derived by
\citeauthor{mashonkina22}\ from their homogeneous NLTE reanalysis
of 10 \sextans\ stars with $-3.2 \leq$ [Fe/H] $\leq -2.6$, 
[Mg/Fe] = $+0.27 \pm 0.08$;
our value is also in agreement with their LTE value,
[Mg/Fe] = $+0.24 \pm 0.08$.
Our mean [Ca/Fe] abundance is lower than than the NLTE derived by
\citeauthor{mashonkina22},
[Ca/Fe] = $+0.31 \pm 0.06$,
but it is in agreement with their LTE value,
[Ca/Fe] = $+0.16 \pm 0.06$.

Previous studies generally
agree that there is a decline in the
[$\alpha$/Fe] ratios at higher metallicities.
There is mild disagreement about the
placement of the knee in the [$\alpha$/Fe] versus [Fe/H] relation.
\citet{reichert20} found a hint that there may be two knees 
in the [Mg/Fe] versus [Fe/H] relation, 
at [Fe/H] $= -2.5$ and $-2.0$,
which could be a consequence of the accretion history of \sextans.
\citet{mashonkina22} discuss this issue in more detail.
Our sample only includes stars with [Fe/H] $< -2.6$,
so we are unable to contribute to this particular debate.

\subsection{Other Light Elements:\ Li, C, N, Na, Al, K}
\label{sec:light}

Li ($Z = 3$) is not detected in any star in our sample.
The upper limits on the Li abundances,
$\log\varepsilon$(Li) $< 0.6$,
are lower than the traditional \citet{spite82} Plateau value,
$\log\varepsilon$(Li) $\approx 2.2$,
and the slight downturn in Li abundances 
found in unevolved stars with [Fe/H] $< -2.8$
\citep{sbordone10}.
The low Li abundances in our stars are consistent with the
well-established phenomenon wherein
Li in the atmospheres is diluted as the
base of the convective zone deepens to hotter layers
during normal stellar evolution up the red giant branch.

C ($Z = 6$) is detected in all five stars in our sample
via the CH A-X (\textit{G}) band.
We derive the C abundance in each star by synthesizing the
CH features in the 4290--4330~\AA\ wavelength region,
using lines from \citet{masseron14}.
The C abundances
have been corrected 
to account for CN processing during normal
stellar evolution \citep{placco14c},
so the values presented in Tables~\ref{tab:abund1}--\ref{tab:abund3}
reflect the natal C abundances.
The corrections for four of the stars are $\approx +0.75$~dex, 
and their corrected [C/Fe] ratios are solar to within a factor of $\approx 2$.
Only one of the five stars, \sexcempno, is C enhanced.
Its evolutionary correction is $+$0.41~dex, yielding a natal
[C/Fe] = $+1.95 \pm 0.20$.
We discuss this carbon-enhanced metal-poor (CEMP)
star in Section~\ref{sec:cempno}.

N ($Z = 7$) is detected only in the CEMP star in our sample
via the CN A-X (red system) bands.
We derive the N abundance from the CN features in the 8000--8100~\AA\
wavelength region, 
using lines from \citet{sneden14}.
The natal N abundance in this star is difficult to infer,
because a wide range of initial---lower---N abundances 
can yield similar surface N abundances
as CN-processed and N-enhanced material 
is dredged up during stellar evolution \citep{placco14c}.
We adopt the current surface abundance,
[N/Fe] $= +1.84 \pm 0.30$, 
as the natal abundance,
but we recommend that it be interpreted with caution.

Na ($Z = 11$) is detected in all five stars in our sample.
The NLTE-corrected [Na/Fe] ratios 
are solar to within a factor of $\approx 2$ in 
four of the five stars.
They fall within the range of metal-poor field stars
and previously examined stars in the inner region of \sextans.
The [Na/Fe] ratio is highly enhanced,
[Na/Fe] = $+1.59 \pm 0.14$, in the CEMP star.

Al ($Z = 13$) is detected in all five stars.
We apply NLTE corrections to the LTE abundances in 
Tables~\ref{tab:abund1}--\ref{tab:abund3}.
Figure~\ref{fig:abundplot}, however, only shows the LTE abundances
for the sake of comparing with literature data, which 
generally have not been corrected for NLTE.~
The [Al/Fe] ratios are within the range of field stars and other
\sextans\ stars.
Two of the stars in our sample exhibit solar [Al/Fe] ratios in NLTE,
which is common among stars with [Fe/H] $< -2$
(e.g., \citealt{andrievsky08,roederer21}).
One star, \mbox{J1018$-$0155},
exhibits significantly enhanced [Al/Fe] = $+0.74 \pm 0.24$ in NLTE.~
Another star, \mbox{J1018$-$0209}, is moderately deficient in Al, with
[Al/Fe] = $-0.40 \pm 0.28$.
No other abundance anomalies are found among light elements
in either of these two stars, and
we lack a satisfactory explanation for the differences in their Al abundances.
There is no reliable Al abundance indicator in our spectrum
of the CEMP star.
The lines of the resonance Al~\textsc{i} doublet at 3944 and 3961~\AA\ 
are detected but heavily blended with CH features.
The high-excitation Al~\textsc{i} doublet at
6696 and 6698~\AA\ is weak and undetected in our spectrum.
We derive an upper limit on the Al abundance in this star,
[Al/Fe] $< +1.72$,
using the latter doublet.

K ($Z = 19$) is detected in all five stars.
K has not been detected previously in any star in \sextans.
The mean NLTE [K/Fe] ratio in the four non-CEMP stars,
$+0.10 \pm 0.04$, falls within the same range as the
mean [$\alpha$/Fe] ratios in these stars.
These [K/Fe] ratios also overlap with those of 
halo stars at similar metallicities.
The CEMP star exhibits highly enhanced K,
[K/Fe] = $+1.28 \pm 0.13$.
This value is higher than that for any star listed in the
JINABase abundance database \citep{abohalima18}.

\subsection{Iron-Group Elements:\ Sc--Zn}
\label{sec:irongroup}

Several iron-group elements, including 
Ti ($Z = 22$), 
V ($Z = 23$), and
Cr ($Z = 24$),
are detected in multiple ionization states.
The differences in the abundances derived from these different states
are generally consistent from one star to another:\
$\approx +0.2$~dex for Ti (with the exception of \sexcempno), 
$\approx +0.3$~dex for V, and
$\approx +0.3$~dex for Cr.
The ions yield higher abundances than the neutrals.
These differences are broadly consistent
with previous NLTE calculations that suggest
the differences in Ti and Cr can be attributed to
NLTE overionization of the minority neutral species
in cool, metal-poor giants
(e.g., \citealt{bergemann10cr,sitnova16}).
Similar NLTE calculations for V have not been made.
The ions should yield more reliable abundances of these species.

The mean [X/Fe] ratios of most iron-group elements are within 
$\approx 0.2$~dex of the solar ratios:
[Sc/Fe] $= -0.21 \pm 0.06$,
[Ti/Fe] $= +0.14 \pm 0.05$,
[V/Fe] $= -0.04 \pm 0.16$,
[Cr/Fe] $= -0.02 \pm 0.12$,
and
[Ni/Fe] $= -0.15 \pm 0.04$.
The mean [Mn/Fe] and [Co/Fe] ratios are
deficient relative to the solar ratios,
$-0.31 \pm 0.05$ and
$-0.33 \pm 0.13$, respectively.
Both Mn and Co are detected only in their neutral states,
which could underestimate their abundances
by several tenths of a dex
(e.g., \citealt{bergemann08,bergemann10}).
The mean [Zn/Fe] ratio is enhanced relative to the solar ratio,
$+0.30 \pm 0.13$.
As shown in Figure~\ref{fig:abundplot},
all of these ratios overlap with the range of ratios in
stars in the inner region of \sextans\ and
metal-poor field stars.

\citet{cowan20} and \citet{sneden23} have shown that the
[Sc/Fe], [Ti/Fe], and [V/Fe] ratios are correlated
in metal-poor field stars.
The mean [Sc/Fe], [Ti/Fe], and [V/Fe] 
ratios in our \sextans\
stars are lower by $\approx$0.1--0.2~dex than the
means in the metal-poor field star samples.
These three ratios
in our \sextans\ stars match the 
low end of the correlations found by \citeauthor{sneden23},
as shown in their Figure~7.
This finding suggests that the supernovae that produced the bulk of the
$\alpha$ and iron-group elements in our \sextans\ stars
were not atypical, yet they produced slight
deficiencies in most elements relative to Fe.
We encourage new theoretical investigations of
supernova yields to better understand this behavior.

\subsection{Heavy Elements:\ Sr--Pb}
\label{sec:heavy}

We detect Sr ($Z = 38$) and Ba ($Z = 56$)
in all stars in our sample, and elements heavier than Ba
can be detected in three of the five stars.
As shown in Figure~\ref{fig:abundplot},
four of the five [Sr/Fe] ratios are comparable to those found
in \sextans\ stars examined previously,
$-0.85 \leq$ [Sr/Fe] $\leq +0.08$.
In contrast, the [Sr/Fe] ratio in the CEMP star,
$-2.37 \pm 0.25$, is $\approx$1~dex lower than
any other star known in \sextans.
The [Ba/Fe] ratios in three of the stars,
$-0.31 \leq$ [Ba/Fe] $\leq -0.10$,
are higher than most other \sextans\ stars with
[Fe/H] $< -2.6$.
The two other stars, including the CEMP star,
exhibit [Ba/Fe] ratios nearly one dex lower.

\begin{figure}
\begin{center}
\includegraphics[angle=0,width=3.35in]{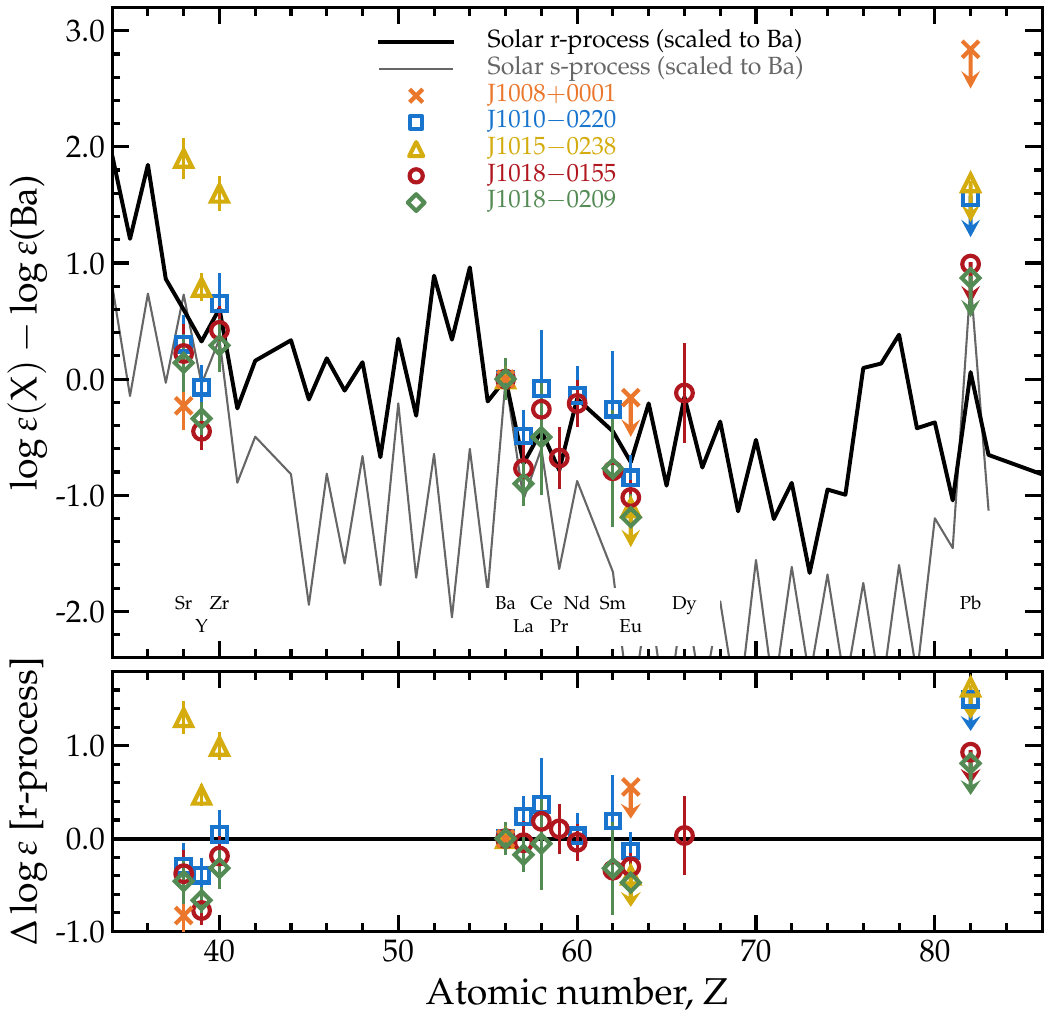}
\end{center}
\caption{
\label{fig:rpropattern}
Heavy-element abundance patterns in the five \sextans\ stars.
The top panel illustrates the abundance patterns,
which have been normalized to
$\log\varepsilon$(Ba) = 0.0.
The bold black line marks the scaled solar \rpro\ pattern, and 
the thin gray line marks the scaled solar \spro\ pattern
\citep{prantzos20}.
The Pb and Bi abundances in the \spro\ pattern have been
enhanced by $+1.0$~dex relative to the solar pattern
to account for the effect of low-metallicity
AGB stars
(cf.\ Figure~20 of \citealt{sneden08},
as calculated in the Appendix of \citealt{roederer10a}).
The bottom panel illustrates the differences between the
observed abundance patterns and the 
scaled solar \rpro\ pattern.
}
\end{figure}

Figure~\ref{fig:rpropattern} illustrates the heavy-element
abundance pattern in the five \sextans\ stars.
The solar system \rpro\ and \spro\ abundance patterns,
normalized to the Ba abundance in each star,
are shown for comparison \citep{prantzos20}.
The \spro\ pattern is disfavored.
Furthermore, enhanced Pb ($Z = 82$) abundances
are also signatures of 
\spro\ enrichment in metal-poor stars
\citep{roederer10c},
and we do not detect an enhanced Pb abundance in any star in our sample.

The abundance patterns in
\mbox{J1010$-$0220}, \mbox{J1018$-$0155}, and \mbox{J1018$-$0209}
are a reasonably close match to the solar \rpro\ pattern.
The most discrepant element, Y ($Z = 39$),
is only discrepant because the solar \rpro\ pattern
overestimates Y by $\approx$0.5~dex
(e.g., \citealt{roederer18c}).
Otherwise, all 11 detected heavy elements lie within
2$\sigma$ of the \rpro\ pattern in these three stars.
Furthermore, the [Ba/Eu] ratio, which is an
indicator of the ratio of \rpro\ to \spro\ material in a star,
is low in these three stars ($-0.80$, $-0.64$, and $-0.47$).
Material where the \rpro\ is dominant will exhibit
[Ba/Eu] $\approx -0.7 \pm 0.2$ 
(e.g., \citealt{sneden08,mashonkina14baeu,prantzos20,roederer23a}),
whereas material where the \spro\ is dominant will exhibit
[Ba/Eu] $> +1$ (e.g., \citealt{sneden08,bisterzo14}).
We conclude that the main component of the \rpro\
is the dominant source of the heavy elements in 
\mbox{J1010$-$0220}, \mbox{J1018$-$0155}, and \mbox{J1018$-$0209}.

Eu ($Z = 63$) is frequently chosen to represent the level of
\rpro\ enhancement in stars.
\mbox{J1010$-$0220}, \mbox{J1018$-$0155}, and \mbox{J1018$-$0209}
are enhanced in \rpro\ elements,
[Eu/Fe] = 
$+0.70 \pm 0.21$,
$+0.33 \pm 0.15$, and
$+0.17 \pm 0.21$, respectively.
\mbox{J1010$-$0220} and \mbox{J1018$-$0155}
are therefore members of the \rone\ class of
moderately \rpro-enhanced stars,
as defined by \citet{beers05} and
revised by \citet{holmbeck20}.
This level of enhancement is not as extreme as found in the
\rpro-enhanced UFD galaxy
\rettwolong\ ($+1.0 <$ [Eu/Fe] $< +2.1$; \citealt{ji16ret2,roederer16b}), 
but it is similar to that in 
the moderately \rpro-enhanced UFD galaxy
\object[NAME TUCANA III]{Tucana~III}
($+0.2 <$ [Eu/Fe] $< +0.6$; \citealt{hansen17tuc3,marshall19}).
Stars with comparable [Eu/Fe] ratios are found in the
\carina, \draco, and \ursaminor\
dSph galaxies,
although only at higher metallicities ([Fe/H] $> -2.5$;
\citealt{shetrone03,cohen09dra,cohen10umi,venn12,norris17b}).

The other two stars in our sample,
\sexcempno\ and \mbox{J1015$-$0238},
exhibit different heavy-element
abundance patterns.
We discuss \sexcempno\ separately in Section~\ref{sec:cempno}.
\mbox{J1015$-$0238} 
has more Sr and less Ba than the other stars in our sample:\
$\log\varepsilon$(Sr/Ba) $= 1.90 \pm 0.23$ 
([Sr/Ba] $= +1.21 \pm 0.25$),
whereas 
$\log\varepsilon$(Sr/Ba) $\approx 0.21 \pm 0.18$
([Sr/Ba] $= -0.48 \pm 0.18$)
for the three \rpro-enhanced stars.
The weak component of the \rpro\ (e.g., \citealt{wanajo13}) and the
weak component of the \spro\
(e.g., \citealt{frischknecht16})
are predicted to be capable of producing enhanced Sr/Ba ratios,
and either process could be responsible
for the heavy elements in \mbox{J1015$-$0238}.
These processes are associated with 
core-collapse supernovae or their progenitor stars.

\subsection{J1008+0001:\ a CEMP-no Star in Sextans}
\label{sec:cempno}

The star \sexcempno\ is located at a projected radius of
10.7~$R_{\rm h}$ (4.3~kpc) from the center of \sextans,
and it is the most widely separated confirmed member of 
\sextans\ at present.
It is highly enhanced
([X/Fe] $> +1.2$) in the light elements
X = C, N, O, Na, Mg, Si, and K.
Its [Ca/Fe] ratio, $+0.50 \pm 0.09$, 
is higher than that found in the other four stars
in our sample, $+0.10 \pm 0.05$.
It is also highly deficient 
([X/Fe] $< -1.4$) in the heavy elements 
X = Sr and Ba.
These characteristics identify \sexcempno\ as a member of the
class of carbon-enhanced metal-poor stars with 
no enhancement of neutron-capture elements
(CEMP-no; \citealt{beers05}).
Such stars are thought to be among the first
Population~II stars to have formed and among the
oldest surviving stars \citep{norris13cemp}.
No CEMP-no stars 
have been identified previously in \sextans.

There have been two measurements of $v_{\rm los}$ of this star.  
One is our measurement ($+223.6 \pm 0.7$~\kmsec; Table~\ref{tab:mike}), 
and the other is an unpublished GIRAFFE measurement 
obtained on 2019 March 12 as part of a separate program 
($+224.4 \pm 5$~\kmsec; S.\ Koposov et al., in preparation).
This star does not exhibit any discernible 
velocity variations over a span of 3~years, 
tentatively suggesting it is not 
part of a binary or multiple star system.

We fit the light-element abundance pattern 
(C through Zn; 6 $\leq Z \leq$ 30)
of \sexcempno\ using the yields predicted for
zero-metallicity Population~III supernovae.
We consider theoretical nucleosynthesis yields
from the grid of 1D supernova models of 
\citet{heger10}, which includes non-rotating stars with initial masses
ranging from 10 to 100~\msun,
explosion energies ranging from 
$0.3 \times 10^{51}$ to $10 \times 10^{51}$~erg,
and various degrees of mixing among the ejecta.
We construct $10^{5}$ representations of the
observed abundance pattern by resampling
the $\log\varepsilon$ abundances from 
Gaussian distributions with standard deviations given by the
observational uncertainties.
We find the best-fit model for each resampled abundance pattern
using a $\chi^{2}$ matching algorithm,
as described in \citet{placco15emp,placco21}.

\begin{figure}
\begin{center}
\includegraphics[angle=0,width=3.35in]{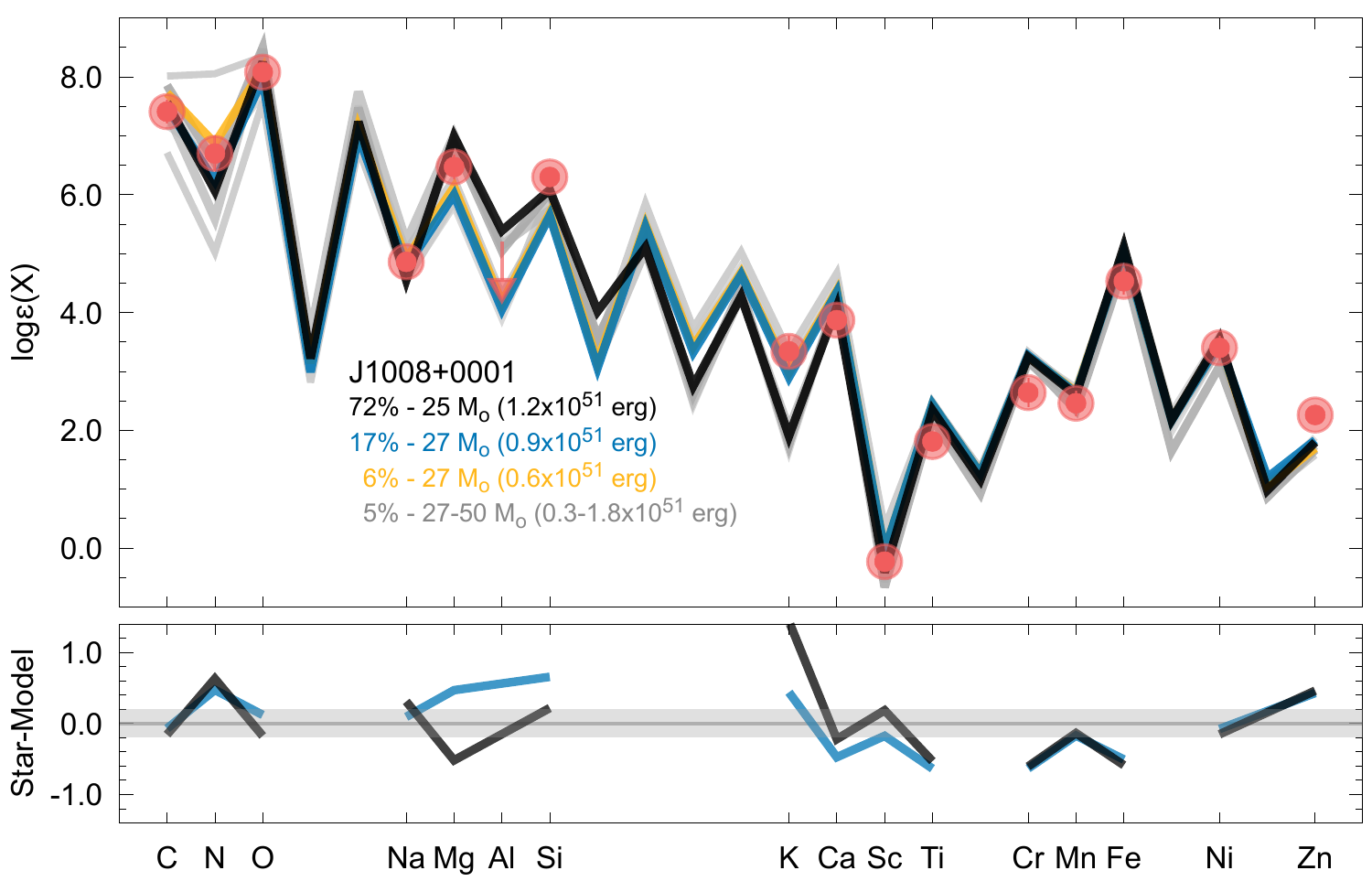}
\end{center}
\caption{
\label{fig:snmodel}
Comparison of the abundances in 
\sexcempno\ with yields predicted by
zero-metallicity supernova models.
The horizontal axis lists the element symbols
at their corresponding atomic number ($6 \leq Z \leq 30$).
Top panel:\
The red dots and downward arrows mark the observed abundance pattern.
The lines mark the predicted nucleosynthesis yields,
and the colors of these lines correspond to the 
properties of the progenitor models indicated in the top panel.
The percentages next to each model 
reflect how often that model was selected as the best fit.
Bottom panel:\
The differences between the observed and best-fit patterns 
are shown for the two models most commonly selected as the best fit.
The horizontal gray line marks a difference of 0.0~dex,
and the shaded gray band marks $\pm$~0.2~dex, an approximate
measure of the 1$\sigma$ observational uncertainties.
}
\end{figure}

Figure~\ref{fig:snmodel} illustrates the results of this test.
We obtain reasonable fits to most elements.
Models with initial masses in the 25--27~\msun\ range
are identified as the best fit $\approx$95\% of the time,
whereas models with initial masses in the 27--50~\msun\ range
are identified $\approx$5\% of the time.
No low-mass models are identified as the best fit for
any realization.
Adopting a N abundance 1~dex lower than the current surface abundance
(Section~\ref{sec:light})
does not appreciably change the distribution of best-fit models.
Our finding, however, must be interpreted with caution.
The chemical evolution models of \citet{hartwig18} predict
that only metal-poor stars with [Mg/C] $< -1.0$ or so
may contain metals produced by a single, dominant progenitor.
The [Mg/C] ratio of \sexcempno\ is $-0.11 \pm 0.24$,
suggesting that it has a low probability of being 
enriched by a single progenitor.
Our results suggest that a massive-star supernova, 
or perhaps a small number of massive-star supernovae,
produced the metals observed today in \sexcempno.

Only two heavy elements are detected in \sexcempno, Sr and Ba.
The Sr/Ba ratio in this star,
$\log\varepsilon$(Sr/Ba) $= -0.23 \pm 0.27$
([Sr/Ba] = $-0.92 \pm 0.32$),
is lower than the other stars in our sample,
all of which contain more Sr than Ba and exhibit
[Sr/Ba] $> -0.55$.
These ratios suggest that the Sr and Ba in \sexcempno\
could have been synthesized by
the weak component of the \spro\ in a 
rapidly rotating low- or zero-metallicity star.
The \citet{frischknecht16} weak \spro\
models predict a wide range of potential [Sr/Ba] ratios,
depending on the conditions found in each star.
These models predict a lower bound in the [Sr/Ba] ratios
of $\approx -0.5$, which is slightly higher than the 
ratio in \sexcempno.
Alternatively, 
an intermediate neutron-capture process (\ipro) operating in
a low- or zero-metallicity massive ($\sim 25$~\msun) star
could also explain the low [Sr/Ba] ratio in \sexcempno\
\citep{banerjee18}.
Either scenario is potentially consistent with the
set of zero-metallicity progenitor models
inferred from the abundances of lighter elements
\citep{roederer14a}.

\section{Discussion}
\label{sec:abundsummary}

\subsection{Heavy Elements in Sextans}

Multiple heavy-element
nucleosynthesis channels were present in the \sextans\ dwarf galaxy.
At least three are apparent among just the five stars in our sample.
One is the main component of the \rpro,
which may occur in neutron-star mergers or exotic 
massive-star supernovae.
The second channel, either the
weak component of the \rpro\ or the weak component of the \spro,
accounts for the enhanced Sr/Ba ratio in
\mbox{J1015$-$0238}.
The third channel, either the
weak component of the \spro\ or the \ipro,
accounts for the deficient Sr/Ba ratio in the
CEMP-no star, \sexcempno.
These three channels can all be associated with
massive-star supernovae or their progenitors.
Finally, previous studies 
\citep{shetrone01,duggan18,theler20}
have detected
material produced by the main component of the
\spro\ in more metal-rich stars
([Fe/H] $> -2.2$)
in the inner regions of \sextans,
representing a fourth
heavy-element synthesis channel.
This channel is associated with low- or intermediate-mass AGB stars.

The abundance ratios produced by these channels 
occupy several distinct regions of chemical space.
Three groups of [Ba/Fe] ratios are found in \sextans:\
one with [Ba/Fe] $\simeq -1$ and $-3.2 <$ [Fe/H] $< -2.3$,
one with [Ba/Fe] $\simeq -0.3$ and $-3.0 <$ [Fe/H] $< -2.7$, and
one with [Ba/Fe] $\simeq +0.3$ and $-2.5 <$ [Fe/H] $< -1.5$, 
as shown in Figure~\ref{fig:abundplot}.
We associate them with the weak \rpro\
(or weak \spro\ or \ipro), 
the main \rpro, and the
\spro, respectively.

Our study has expanded the range of 
heavy-element enrichment processes known to occur in \sextans.
Nevertheless,
several \sextans\ stars still lack sufficient chemical information
to reliably diagnose the nucleosynthetic origin(s) 
of their heavy elements.
Followup observations are warranted
to better understand which scenarios
occurred in \sextans.

\subsection{Chemical Inhomogeneity in Sextans}

The chemical diversity among the five stars in our sample
suggests that
stars in the outskirts of \sextans\ formed in
chemically inhomogeneous regions.
In contrast, stars in the inner region of \sextans\
are more chemically homogeneous among the
$\alpha$ and \ncap\ elements at a given metallicity
\citep{aoki20,lucchesi20,theler20}.
The \sextans\ dSph also contrasts with the three UFD galaxies
studied by \citet{waller23},
who found that stars in their outer regions
were chemically similar to those near their centers.

Very few CEMP-no stars have been confirmed
among stars studied in dSph galaxies:\
two stars in \carina\ \citep{susmitha17,hansen23carina},
one star in \draco\ \citep{cohen09dra},
two stars in \sculptor\ \citep{skuladottir15,skuladottir23scl}, 
two stars in \ursaminor\ \citep{cohen10umi},
and
possibly one star in \canvenone\ \citep{yoon20}.
The dSph galaxies contrast with the UFD galaxies
in this regard, because the
occurrence frequency of CEMP-no stars in 
UFD galaxies is relatively high
(e.g., \citealt{norris10seg1,frebel14seg1,spite18cempno,ji20,chiti23}).
On the other hand, 
a focused study by \citet{chiti18cemp} revealed that 
the CEMP fraction among stars with [Fe/H] $< -3.0$ in 
the \sculptor\ dSph, $36 \pm 8$\%,
is not different from that of the Milky Way halo, $\approx 42$\%.
\citeauthor{chiti18cemp}\
noted, however, that none of the CEMP stars in their \sculptor\ sample
exhibited [C/Fe] $> +1$.
That property is different from the
Milky Way halo and UFD galaxies,
where stars with [C/Fe] $> +1$ are more common.
\citet{skuladottir23scl} reached a different conclusion
from their sample of 11 stars in \sculptor\ with
[Fe/H] $< -3.0$, finding only one CEMP-no star.
A fresh analysis may be necessary to resolve
this apparent discrepancy in the \sculptor\ dSph.

Our study is not equipped to derive the
CEMP fraction in \sextans.
Our results
suggest that the outer regions of \sextans,
and by extension other more massive dSph galaxies,
could be reservoirs of extreme
CEMP-no stars. 
One possible scenario is that
these regions may have been similar 
to those where lower-mass galaxies formed,
thereby establishing a common chemical enrichment pathway
between galaxies of differing masses.
Another possible scenario is that
these star-forming 
regions could have been actual UFD galaxies.
This idea is supported by the recent work of \citet{deason23},
who found that the stellar metallicity distribution of \sextans\
could allow for the accretion of multiple UFD-like systems.
Much larger samples of stars at large radius will be necessary 
to distinguish among these scenarios.

\subsection{Substructure in Sextans}

Recent observations suggest that
extended stellar halos may be a relatively common feature
of dwarf galaxies
(e.g., \citealt{chiti21,stringer21,yang22,sestito23umi}),
including \sextans\ \citep{qi22},
even in the absence of tidal distortions.
The stars in our sample are located at much larger radii than the
stellar substructures in \sextans\ identified by previous work
using stellar velocities and metallicities.
We thus cannot directly associate the stars in our sample
with that substructure.
Future studies of larger samples of stars
at large radius 
will be necessary to potentially associate
these chemical signatures with
dynamical substructures in \sextans.

\section{Conclusions}
\label{sec:conclusions}

We have collected high-resolution, high-S/N optical spectra
of five confirmed member stars of the \sextans\ dSph galaxy that
are located at projected distances of 
3.5 to 10.7 $R_{\rm h}$
(1.4 to 4.3~kpc)
from its center.
We identify several chemical signatures
absent from previous samples of \sextans\ stars,
including CEMP-no, \rpro, and enhanced Sr/Ba abundance signatures.

Our results indicate that production of the lighter elements,
including $\alpha$, odd-$Z$, and iron-group elements,
was dominated by core-collapse supernovae
at early times.
The mildly enhanced [$\alpha$/Fe] ratios,
which are lower in our sample than in typical metal-poor field stars,
could indicate a deficiency of metals produced by the
highest-mass stars.
The outskirts of dSph galaxies, such as \sextans,
could represent one birth environment for 
metal-poor stars occupying the low end of the distribution of
[Sc/Fe], [Ti/Fe], and [V/Fe] ratios
identified by \citet{cowan20}.
Three stars exhibit moderate enhancement of \rpro\ elements.
One CEMP-no star exhibits evidence of enrichment 
dominated by a supernova that produced a chemical 
signature distinct from that found in the other four stars.
All of these chemical signatures can be attributed to
enrichment from
low-metallicity massive stars, their supernovae,
or mergers of neutron stars that result from 
such supernovae.

We conclude that at least some of the stars in our sample formed
in regions with different chemical evolution histories
than the stars at the center of \sextans.
We anticipate that future studies of 
stars at large radius in \sextans\
and other dSph galaxies
will reveal a rich diversity of chemical signatures
from the first generations of stars and supernovae.

\acknowledgments

We thank the referee for their helpful comments that have 
improved this manuscript.
We thank the dedicated staff at Las Campanas Observatory
for their efforts to develop remote observing
capabilities during the COVID-19 travel restrictions.
EO wishes to remember Jill Bechtold here.
We acknowledge generous support 
from the U.S.\ National Science Foundation (NSF),
provided by grants 
AST-1815403 and AST-2205847 (IUR, MM);
AST-1813881, AST-1909584, and AST-2206046 (ABP, SK, MGW);
AST-1812461 (NC);
AST-1815767 (EWO);
and 
PHYS-1430152 (Physics Frontier Center/JINA-CEE; IUR).~
The work of VMP is supported by NOIRLab, which is
managed by AURA under a cooperative agreement with the NSF.~
This research has made use of NASA's
Astrophysics Data System Bibliographic Services;
the arXiv pre-print server operated by Cornell University;
the SIMBAD and VizieR
databases hosted by the
Strasbourg Astronomical Data Center;
the ASD hosted by NIST;
and the
IRAF software package, 
which was distributed by the National Optical Astronomy Observatory,
which was managed by AURA
under a cooperative agreement with the NSF.~

\facility{%
Magellan (MIKE),
MMT (Hectochelle),
VLT (FLAMES),
}

\software{%
IRAF \citep{tody93},
matplotlib \citep{hunter07},
MOOG \citep{sneden73},
numpy \citep{vanderwalt11},
R \citep{rsoftware},
scipy \citep{jones01}
}

\bibliographystyle{aasjournal}
%\bibliography{../../iuroederer}
\bibliography{ms.bbl}

\end{document}